\documentclass[12pt, draftclsnofoot, onecolumn]{IEEEtran}

\ifCLASSINFOpdf
\else

\usepackage[dvips]{graphicx}
\usepackage[usenames,dvipsnames]{pstricks}
\usepackage{mathrsfs}
\usepackage{amsmath}
\usepackage{times}
\usepackage[mathscr]{euscript}
\usepackage{subfig}

\newlength\figureheight
\newlength\figurewidth

\usepackage{times}
\usepackage{epsfig}
\usepackage{microtype}
\usepackage{caption}
\usepackage{amsbsy}

\usepackage[linesnumbered,ruled,vlined]{algorithm2e}

\usepackage{amsthm}
\theoremstyle{definition}

\RequirePackage{amssymb}

  \graphicspath{{../eps/}}
   \DeclareGraphicsExtensions{.eps}
  \DeclareGraphicsExtensions{.pdf,.jpeg,.png}
\fi

\usepackage{lipsum}

\begin{document}


\title{Two-Level Spatial Multiplexing using \\
Hybrid Beamforming Antenna Arrays for \\mmWave
Communications
}

\author{\large \IEEEauthorblockN{Xiaohang Song, Nithin Babu, Wolfgang Rave,\\ Sudhan Majhi,
\textit{IEEE Senior Member}, and Gerhard Fettweis, \textit{IEEE Fellow}} \vspace*{-3mm}
\thanks{This work has been supported by the priority program SPP 1655 "Wireless Ultra High Data Rate
Communication for Mobile Internet Access" by the German Science Foundation (DFG).

X. Song, W. Rave, and G. Fettweis are with Vodafone Chair, Technische Universit\"at
Dresden, Dresden, Germany, e-mail: \{xiaohang.song, wolfgang.rave,
gerhard.fettweis\}@tu-dresden.de. N. Babu, and S. Majhi are with Indian Institute of
Technology, Patna, India, e-mail: \{nithin.mtcm14, smajhi\}@iitp.ac.in. } }

\maketitle

\begin{abstract}
In this work, we consider a two-level hierarchical MIMO antenna array system, where each antenna of the upper level is made up of a subarray on the lower one. The concept of spatial multiplexing is applied twice in this situation: Firstly, the spatial multiplexing of a Line-of-Sight (LoS) MIMO system is exploited. It is based on appropriate (sub-)array distances and achieves multiplexing gain due to phase differences among the signals at the receive (sub-)arrays. Secondly, one or more additional reflected paths of different angles (separated from the LoS path by different spatial beams at the subarrays) are used to exploit spatial multiplexing between paths. 

By exploiting the above two multiplexing kinds simultaneously, a high dimensional system with maximum spatial multiplexing is proposed by jointly using 'phase differences' within paths and 'angular differences' between paths. The system includes an advanced hybrid beamforming architecture with large subarray separation, which could occur in millimeter wave backhaul scenarios. The possible gains of the system w.r.t. a pure LOS MIMO system are illustrated by evaluating the capacities with total transmit power constraints.
\end{abstract}



%
\IEEEpeerreviewmaketitle

\section{Introduction}

Around 2020 peak data rates in cellular networks are expected to be in the order of
10~$\mathrm{Gb/s}$ \cite{GF_2020}. Base stations will serve multiple sectors
\cite{Marzetta_massiveMIMO} and will be no more than 100~$\mathrm{m}$ apart in urban
areas. Our previous work \cite{Song_GC2015} showed great potential in building ultra
high speed fixed wireless links to meet this growing demand for high capacity of the
front/back-haul over a \emph{single} LoS path. For future dense networks, wireless front-
and/or backhaul links offer easy and cheap deployment in comparison with costly optical
fibers. The unlicensed 60~$\mathrm{GHz}$ band has become the most popular for this purpose
due to large available bandwidth, high frequency reuse and reasonable array sizes which
could fully exploit the spatial multiplexing gains in LoS MIMO channels.

The works in \cite{Larsson, Haustein2003} derived optimal antenna arrangements on
parallel planes in terms of antenna/subarray distances that provide self-orthogonal LoS
channel matrices. However, the same kind of spatial multiplexing remains possible for
antenna arrangements on tilted non-parallel planes \cite{BohagenConstructionCapacity,
BohagenOptimalDesignUPA, ChunhuiZhou, Song_2015} or for even more complicated 3D
arrangements \cite{Song_2015}.

Our work is motivated by the potential of having higher capacities, if additional paths
that occur under some oblique angles w.~r.~t. the LoS direction become available and can
be discriminated using beamforming. Ref.~\cite{Larsson} showed high robustness of the
spatial multiplexing gain in LoS MIMO against displacements like translation and
rotation. Therefore, the optimal geometrical arrangements need not to be realized with
high accuracy and a significant multiplexing gain can still be expected using a
reflected path with large antenna (rather subarray) separation. In this way, we will
establish a link between two spatial multiplexing approaches under LoS conditions
\cite{Larsson} and under multipath conditions as originally envisaged
by \cite{Telatar99, Foschini1996}.


Large numbers of closely packed antennas are normally demanded by mmWave systems for compensating high attenuation. This does not allow one RF chain per antenna element, due to hardware cost, power and space constraints. Thus a hybrid architecture, jointly using analog beamforming in the RF frontend and digital beamforming in baseband processing, is of our interest. Differently superposed analog signals are down-converted to baseband and create a set of spatial streams. Such hybrid beamforming techniques provide greater implementation flexibility in comparison to fixed analog solutions and lower hardware cost in comparison to fully digital solutions \cite{Hybrid_switching_05, Veen_Hybrid_10, Heath_Hybrid_12}. Antenna selection concepts \cite{Hybrid_switching_05} that rely on custom RF switch networks can provide an additional degree of freedom for shaping the beam patterns and directions.


The rest of this paper is organized as follows: In Section~\ref{sec:systemModel}, we
present channel and system models which exploit two spatial multiplexing kinds separately. At first, we
consider multiplexing over a \textit{single} LOS path (Sec.~\ref{subsec:LOS_MIMO}). This is
contrasted with a limited scattering environment for which a multiplexing gain over
\textit{multiple} paths is obtained in Sec.~\ref{sec:Model_single_subarray}. In view of
the intended application in mmWave communications, a description in terms of a hybrid
beamforming architecture using a set of available analog beam patterns is presented here.
After considering these limited cases, we propose a transmission model which combines
the approaches and exploits the above two kinds of spatial multiplexing jointly in
Sec.~\ref{sec:two_level_MIMO}. Section~\ref{sec:spectral_Efficiency_Evaluation}
proposes the spectral efficiency under a sum power constraint as the benchmark for the spatial multiplexing gain in our two-level multiplexing scenario. The optimization problem is converted to a power allocation problem and the solution can be given by waterfilling algorithm. Numerical results and a discussion are presented in Section~\ref{sec:Results_and_Discussion} before we summarize our work in
Section~\ref{sec:conclusion}.

\textit{Notation:} Upper- and lowercase variables written in boldface, such as
$\mathbf{A}$ and $\mathbf{a}$, denote matrices and vectors; $a$ in normal font refers to a scalar; $(\cdot)^{\mathrm{T}}$, $(\cdot)^{\mathrm{H}}$ and $||\cdot||_{F}$ denote transpose, conjugate transpose and Frobenius norm, respectively; $\mathrm{tr}(\cdot)$ and $\mathrm{det}(\cdot)$ denote the trace and the determinant, respectively; $\mathbf{A} \otimes \mathbf{B}$ is the Kronecker product of
$\mathbf{A}$ and $\mathbf{B}$; $\{\mathbf{A}\}_{lk}$ denotes element $(l,\ k)$ of
$\mathbf{A}$ and $|a|$ denotes the absolute value of $a$. Expectation is denoted by $\mathbb{E}[\cdot]$ and
$\mathbf{I}_N$ is the $N \times N$ identity matrix; $\mathcal{CN}(\mathbf{a}, \mathbf{A})$ is a complex Gaussian random vector with mean $\mathbf{a}$ and covariance matrix $\mathbf{A}$.


\section{Single- vs. Multi-path Spatial Multiplexing}
\label{sec:systemModel}

\begin{figure}[ht]
\centering
\subfloat[]{
   \includegraphics[width=0.49\textwidth]{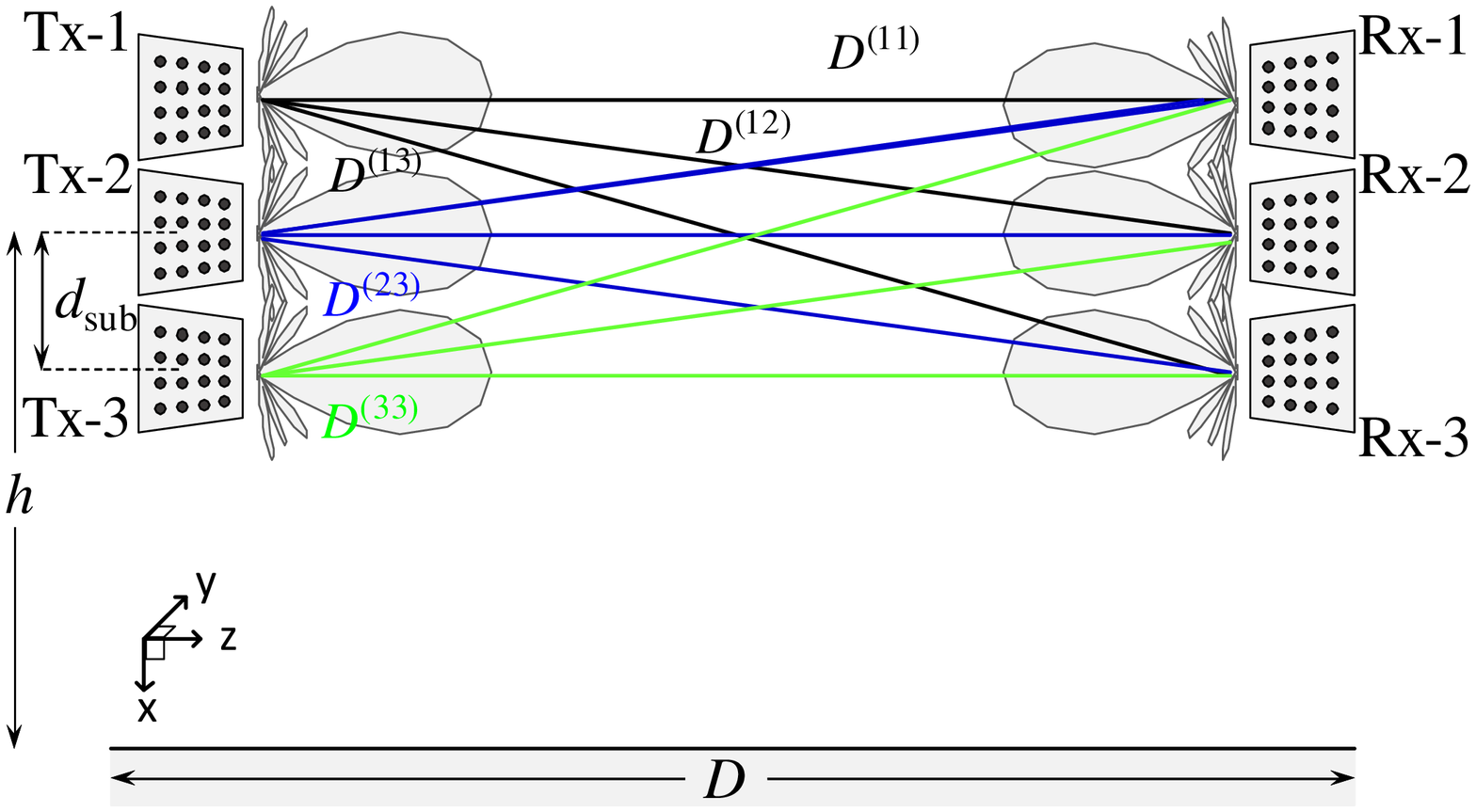}
}
\subfloat[]{
   \includegraphics[width=0.49\textwidth]{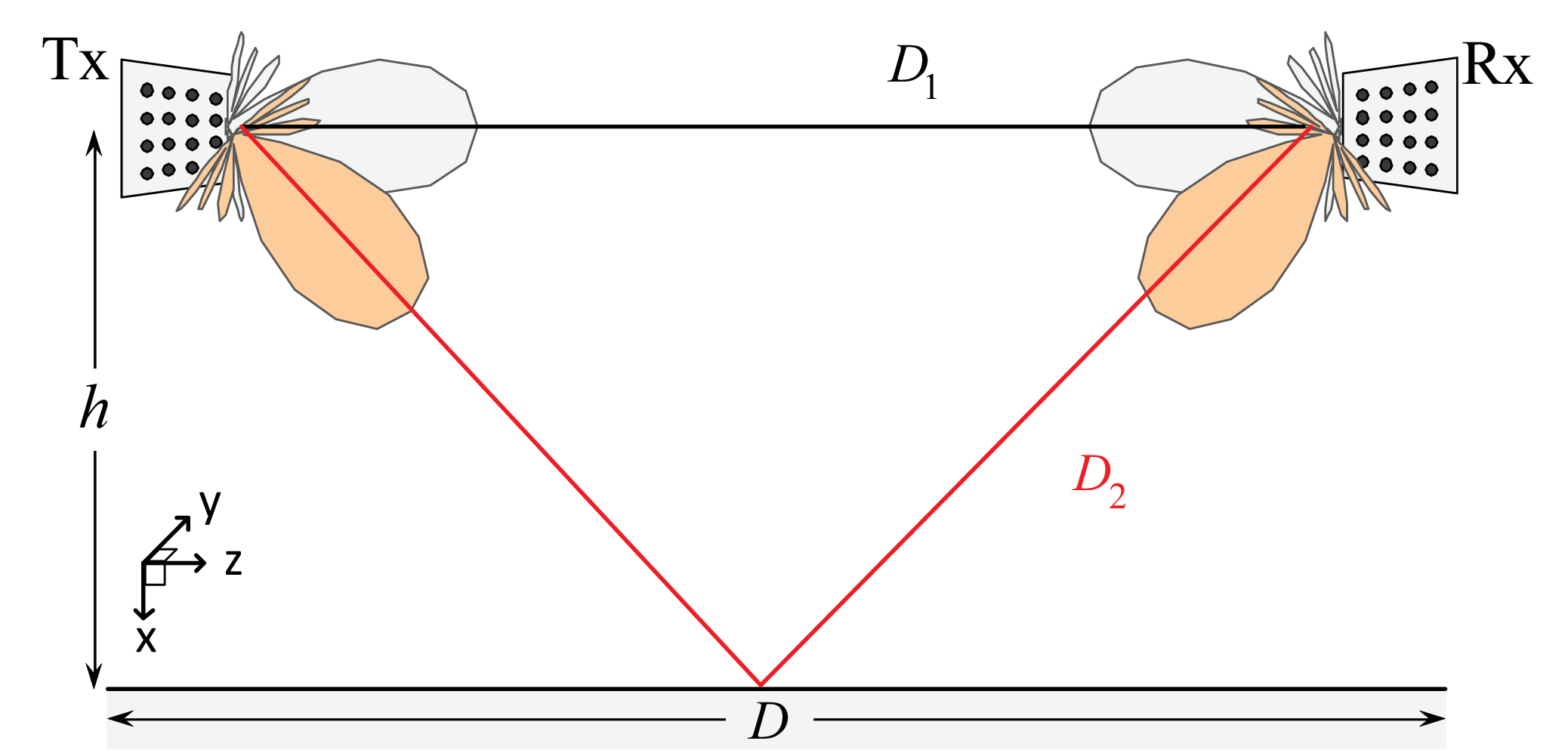}
}\\\vspace{-3mm} 
\subfloat[]{
   \includegraphics[width=0.49\textwidth]{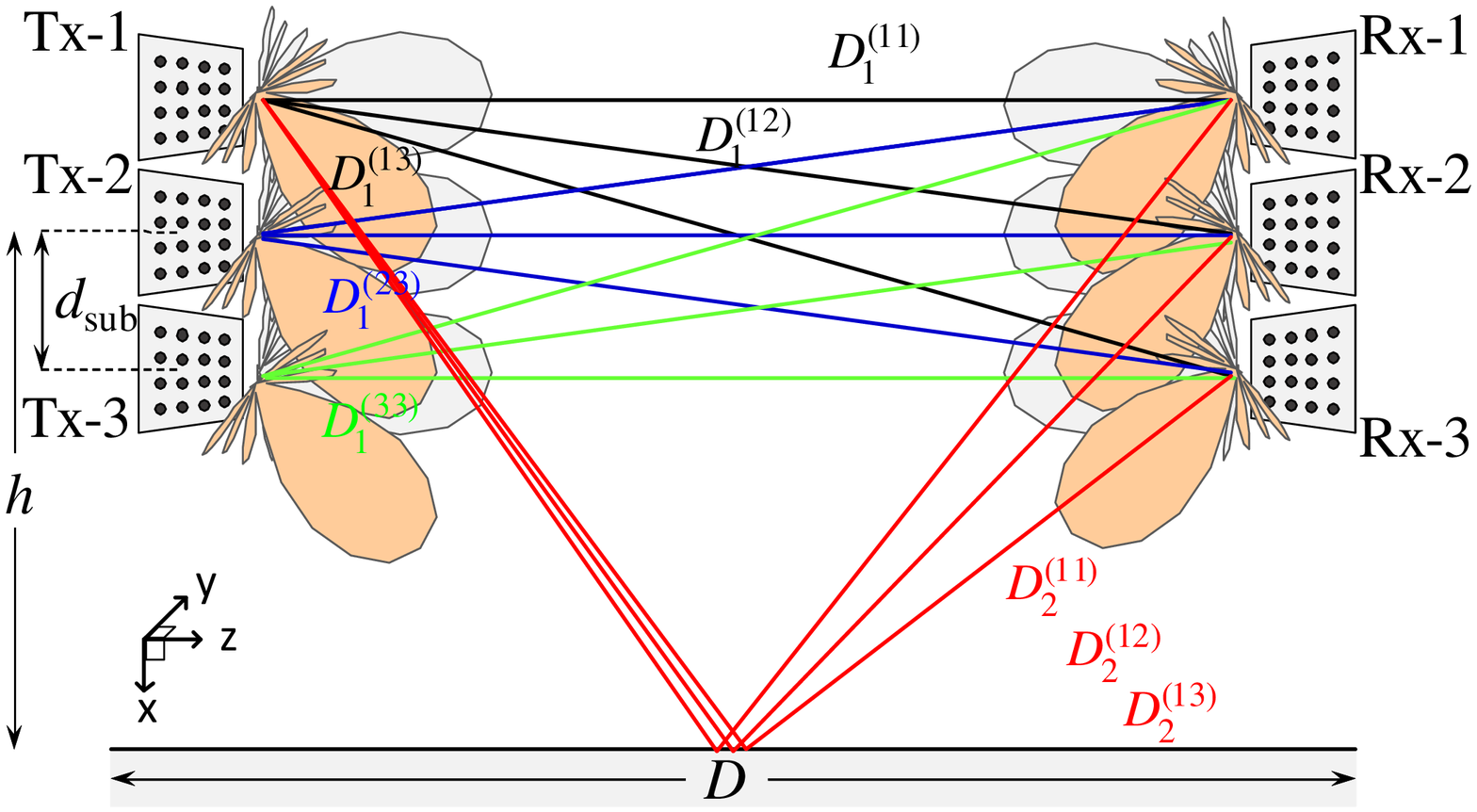}
}
\center \vspace{-7mm} \caption{\small Geometry for (a) LOS spatial multiplexing between
subarrays $(\max\,\{N_\mathrm{s}\}=3)$, (b) multipath spatial multiplexing
between single subarrays over two paths $(\max\,\{N_\mathrm{s}\}=2)$, and (c) the simplest example of two-level
spatial multiplexing: over each of the two paths (LOS connection and ground reflection),
three streams are multiplexed between subarrays at transmitter and receiver
side $(\max\,\{N_\mathrm{s}\}=6$, s.t. $\lambda \ll d_\mathrm{sub} \ll h<D)$} \vspace{-5mm} \label{fig:system_geometry_relations}
\end{figure}

In this section, we present detailed transmission models for two spatial multiplexing kinds, namely 'spatial multiplexing over a single path' (usually the LoS path) and 'multiplexing of spatial streams in a multipath scenario'. In the first case (see Fig.~\ref{fig:system_geometry_relations}(a)), a description in terms of an array of subarrays is considered, as the subarrays provide necessary antenna gain in mmWave links for LoS MIMO communication. The second case can be viewed as the '\textit{conventional}' way of spatial multiplexing~\cite{Foschini1996}. It is shown schematically in Fig.~\ref{fig:system_geometry_relations}(b), where we assume initially only a single antenna array at transmitter (Tx) and receiver (Rx) side. Signals traveling along different paths/directions are addressed with beam steering algorithms. These two cases summarize the state-of-the-art works that exploit two kinds of spatial multiplexing separately. In next section, we merge these two approaches into a two-level hierarchical MIMO system with appropriately large subarray separation to exploit both kinds of spatial multiplexing gains simultaneously (see Fig.~\ref{fig:system_geometry_relations}(c)).

\subsection{Spatial Multiplexing under Line-of-Sight Conditions}
\label{subsec:LOS_MIMO}

The essential insights to achieve spatial multiplexing between antenna arrays over a
\textit{single} path were developed originally for line-of-sight MIMO
communication~\cite{Larsson} using a carrier frequency $f$ with corresponding wave
length $\lambda=c/f$. Let us also consider the geometry of the situation as sketched in
Fig.~\ref{fig:system_geometry_relations}(a) and describe it with a spherical-wave
model\footnote{As shown in \cite{Pu2015}, the spherical-wave model is more accurate and
leads to larger spectral efficiency of the links than the conventionally used plane-wave model, if the antenna separation
$d_\mathrm{sub}$ is in the order of $\sqrt{\lambda D}$.}. Let us assume that transmitter and receiver are separated by a link distance $D$ and each side consists of $N$ subarrays/antennas\footnote{More precisely, we will denote the effective propagation length along path $p$ between the $l$-th transmit subarrays/antennas and the $k$-th receive subarrays/antennas as $D_p^{(lk)}$ later in this work.}. This would form a standard LoS MIMO scenario, and the spectral efficiency depends on the spacing $d_\mathrm{sub}$ between the subarrays/antennas in a 'super' array. The optimal spacing $d_\mathrm{sub}$ to a 'super' array of $N$ elements is provided by~\cite{Larsson}
\begin{equation}
   d_\mathrm{sub} = \sqrt{\frac{\lambda D}{N}} \, ,
\end{equation}
which relies on the relation
\begin{equation}
   \lambda\ll d_\mathrm{sub}\ll D.
   \label{equ:distance_relation}
\end{equation}

If this condition is fulfilled, the propagation distances between different pairs of subarrays (antennas) are negligible when one calculates the path attenuation values. However, while pathloss differences can be neglected, the length of the \emph{very same} propagation paths between transmit and receive antennas will differ by certain fractions of $\lambda$. These differences provide specific phase shifts between the observed signals at the receive subarrays/antennas. As a consequence of the conditions stated by
Equ.~\eqref{equ:distance_relation}, the resulting channel or '\textit{coupling}' matrix
$\mathbf{H}_{\mathrm{LoS}}$ between transceiver arrays can be optimized to obtain a
spatially orthogonal matrix with $\mathbf{H}_{\mathrm{LoS}}^\mathrm{H} \,
\mathbf{H}_{\mathrm{LoS}} = N \cdot \mathbf{I}_{N}$.

Let us illustrate this scenario by an example: considering the entries of $\mathbf{H}_{\mathrm{LoS}}$ in a symmetric system with two uniform linear arrays (ULAs). The two arrays consist of $N$ subarrays each and are arranged in two parallel lines. Both lines are perpendicular to the transmit direction and the radiated signals are traveling in a free space. The matrix elements representing the phase coupling between subarrays at different sides can then be written as $\{\mathbf{H}_{\mathrm{LoS}}\}_{lk}= e^{-j\frac{2\pi}{\lambda}\cdot D^{(lk)}} \approx e^{-j2\pi D/\lambda}\cdot e^{-j\pi (l-k)^2/N}$, where $D^{(lk)}$ is the distance between the $l$-th transmit and $k$-th receive subarray/antenna. Specifically, for $N=3$ we find
\begin{equation}
\mathbf{H}_{\mathrm{LoS}} = \left [ \begin{array}{ccc}
   1                  & e^{-j\frac{\pi}{3}}  &  e^{-j\frac{4\pi}{3}} \\
 e^{-j\frac{\pi}{3}}  & 1                    &  e^{-j\frac{\pi}{3}} \\
 e^{-j\frac{4\pi}{3}} & e^{-j\frac{\pi}{3}}  & 1
\end{array} \right]\cdot e^{-j2\pi D/\lambda}.
\label{eq:coupling_example}
\end{equation}

The baseband model describing the transmission of $N$ data streams between the subarrays/antennas can be expressed in the form of a simple linear model as
\begin{equation}
\mathbf{y} = \rho \cdot \mathbf{H}_{\mathrm{LoS}} \, \mathbf{s} + \mathbf{n} \, ,
\label{equ:LOS_system}
\end{equation}
where $\mathbf{s}$, $\mathbf{y}$ are transmit symbol vector and receive vector of the size $N \times 1$. $\rho$ indicates the common
channel gain between the subarrays/antennas (including array/antenna gains and pathloss). $\mathbf{n}$ is i.i.d. zero mean complex white Gaussian noise distributed as $\mathbf{n}\sim \mathcal{CN}(\mathbf{0}, \sigma^2_{n}\cdot\mathbf{I}_{N})$. The LoS system supports $N_\mathrm{s}=N$ data streams simultaneously.

Extending this idea, orthogonality can also be achieved when uniform rectangular arrays (URAs) or uniform square arrays (USAs) are used. Assuming that each column of the array has $N_\mathrm{x}$ subarrays/antennas along the $\mathrm{x}$-axis and each row has $N_\mathrm{y}$ subarrays/antennas along the $\mathrm{y}$-axis, the transceiver arrays would consist of $N=N_\mathrm{x}\cdot N_\mathrm{y}$ elements each. The phase coupling matrix $\mathbf{H}_{\mathrm{LoS}}$ can then be factorized into a Kronecker product of two phase coupling matrices of ULAs along orthogonal directions \cite{Larsson} as
\begin{equation}
   \mathbf{H}_{\mathrm{LoS}}= \mathbf{H}_{\mathrm{LoS},\mathrm{x}} \otimes
   \mathbf{H}_{\mathrm{LoS}, \mathrm{y}} \, ,
   \label{equ:URA_factorization}
\end{equation}
where $\mathbf{H}_{\mathrm{LoS},\mathrm{x}}$ and $\mathbf{H}_{\mathrm{LoS},\mathrm{y}}$ denote two phase coupling matrices of ULAs with $N_\mathrm{x}$ and $N_\mathrm{y}$ elements, respectively. In case that both the $\mathrm{x}$-axis and the $\mathrm{y}$-axis arrays satisfy the optimal ULA arrangements, $\mathbf{H}_{\mathrm{LoS},\mathrm{x}}$ and $\mathbf{H}_{\mathrm{LoS},\mathrm{y}}$ are then obviously orthogonal matrices as before. Therefore, it still holds that $\mathbf{H}_{\mathrm{LoS}}^\mathrm{H} \cdot \mathbf{H}_{\mathrm{LoS}}=N\cdot \mathbf{I}_N$. In later numerical evaluations, we limit the subsequent treatment to the case of ULAs of subarrays along the $\mathrm{x}$-axis, while the power gain at each subarrays are achieved by 2D arrays for reasonable link budgets. Meanwhile, the direction of $\mathrm{x}$-axis is assumed to be perpendicular to the ground for simplicity.

\subsection{Spatial Multiplexing under Multipath Conditions}
\label{sec:Model_single_subarray}

Spatial multiplexing gain in multipath scenario was originally studied in the seminal works by \textit{Telatar}, \textit{Foschini} et.al. \cite{Telatar99, Foschini1996} with a rich scattering environment. In the context of mmWave communications, the environment is assumed to be of limited scattering, and the scenario is addressed using hybrid beamforming with limited signal processing complexity and power consumption \cite{Heath_Hybrid_12}. Thus we assume that $P$ paths\footnote{It is interesting to note that for a single path under the assumption of planar waves, no spatial multiplexing gain is possible.} with signals from \emph{different} directions are available (see Fig.~\ref{fig:system_geometry_relations}(b), where we schematically illustrate the simplest possible situation of two paths).

\begin{figure}[t]
\centering
   \includegraphics[width=\textwidth]{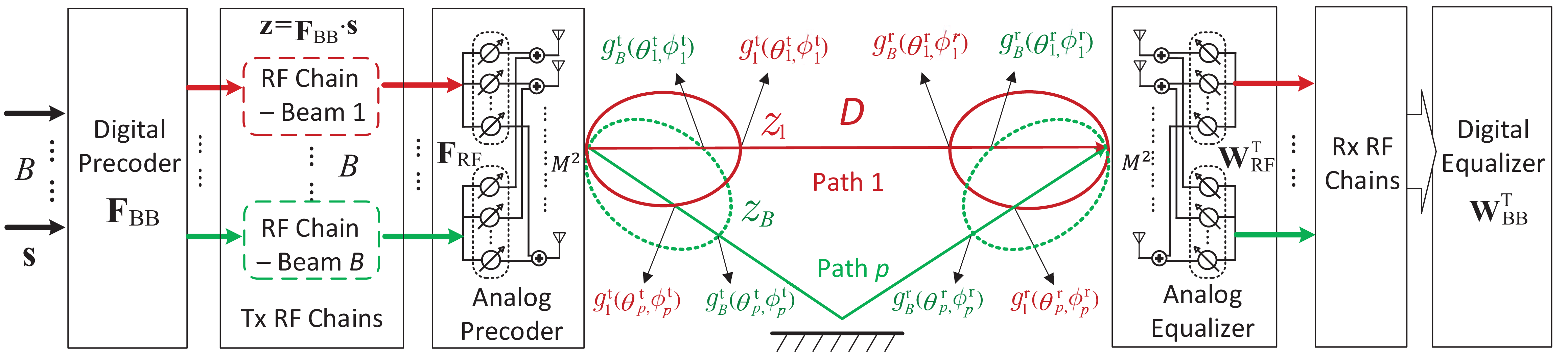}
\center \vspace{-2mm} \caption{\small Simplified block diagram of a mmWave single subarray system with a conventional hybrid beamforming architecture \cite{Heath_Hybrid_12}.} \label{fig:system_block_diagram_B}
\vspace{-6mm}
\end{figure}

As a complementary approach to obtain spatial multiplexing gain, we firstly assume only one single USA at transmitter side and one single USA at receiver side, both at height $h$. The arrays are facing each other over a link distance $D$. Each of two arrays consists of $M\times M$ antenna elements modeled as isotropically radiating point sources (later such a USA will be viewed as one of several subarrays).

In the general cases, $P$ reflecting objects may exist for $P$ paths, and each reflection object is assumed to contribute a single propagation path $p$. The direction of the path $p$ is characterized by its elevation and azimuthal angles of departure and arrival, respectively. These angles are denoted as $\{
\theta^{\mathrm{t}}_{p}, \, \phi^{\mathrm{t}}_{p} \}$ as well as $\{
\theta^{\mathrm{r}}_{p}, \, \phi^{\mathrm{r}}_{p} \}$ for path $p$. The superscripts $\{\mathrm{t}$, $\mathrm{r}\}$ indicate transmitter and receiver, respectively.

Considering the hardware constraints at subarrays, we assume that a subarray is supported by $B$ RF chains. Therefore, maximum $B$ beams are radiated by the analog beamforming algorithms towards the $P$ available paths and may support $N_{\mathrm{s}}$ ($N_{\mathrm{s}} \le P$, $N_{\mathrm{s}} \le B$) data streams. We write the transmission model in digital baseband as
\begin{equation}
\mathbf{y} = \mathbf{W}_{\mathrm{BB}}^{\mathrm{T}}
\underbrace{\mathbf{W}_{\mathrm{RF}}^{\mathrm{T}}\, \mathbf{H}\,
\mathbf{F}_{\mathrm{RF}}}_{\triangleq \mathbf{H}_{\mathrm{eff}}} \mathbf{F}_{\mathrm{BB}}\, \mathbf{s} +\mathbf{W}_{\mathrm{BB}}^{\mathrm{T}} \underbrace{\mathbf{W}_{\mathrm{RF}}^{\mathrm{T}} \, \mathbf{n}}_{\mathbf{n}_{\mathrm{eff}}} \, , \label{equ:SISO_system}
\end{equation}
where the matrix products $\mathbf{W}_{\mathrm{BB}}^{\mathrm{T}}\mathbf{W}_{\mathrm{RF}}^{\mathrm{T}}$ and $\mathbf{F}_{\mathrm{RF}}\mathbf{F}_{\mathrm{BB}}$ reflect the hybrid beamforming approach. The vectors $\mathbf{s}$ and $\mathbf{y}$ of size
$B \times 1$ denote the transmit and receive symbols. $\mathbf{n}$ corresponds to zero mean complex white Gaussian noise distributed as $\mathbf{n}\sim \mathcal{CN}(\mathbf{0}, \sigma^2_{\mathrm{n}}\cdot \mathbf{I}_{M^2})$. The $B \times B$ matrices $\mathbf{F}_{\mathrm{BB}}$ and $\mathbf{W}_{\mathrm{BB}}$ act as baseband precoder and equalizer, respectively. As shown in the numerical evaluation later, when maximizing spectral efficiency with different power constraints in different scenarios, the number of supported data streams may change. Therefore, in order to select and support $N_{\mathrm{s}}$ data streams from $\mathbf{s}$, $N_{\mathrm{s}}\le B$, one may expect only first $N_{\mathrm{s}}$ columns of $\mathbf{F}_{\mathrm{BB}}$ and $\mathbf{W}_{\mathrm{BB}}$ contain non-zero values. 

The matrices $\mathbf{F}_{\mathrm{RF}}\in{\mathbb{C}^{M^2\times B}}$, $\mathbf{W}_{\mathrm{RF}}\in{\mathbb{C}^{M^2\times B}}$ contain $B$ beam patterns in their columns, which are realized simultaneously with \emph{analog} beamforming at RF front-ends (see Fig.~\ref{fig:system_block_diagram_B}). Typically, the entries in each column are of constant magnitude. Meanwhile, they provide phase shifts between corresponding baseband signals and pass-band signals at antenna elements (implemented with e.g., a set of
phased arrays \cite{Mailloux2005}). The superposed pass-band signals are sent via a set of beams with different steering directions. The matrices $\mathbf{F}_{\mathrm{BB}}$, $\mathbf{F}_{\mathrm{RF}}$, $\mathbf{W}_{\mathrm{BB}}$ and $\mathbf{W}_{\mathrm{RF}}$ allow joint \emph{digital} and \emph{analog} hybrid beamforming and are optimized according to some criterion, e.g., maximizing spectrum efficiency. The actually transmitted baseband symbol vector after baseband processing is defined as $\mathbf{z}\triangleq\mathbf{F}_{\mathrm{BB}} \, \mathbf{s}$. Meanwhile, the actually radiated symbol vector on pass-band is $\mathbf{x}\triangleq\mathbf{F}_{\mathrm{RF}} \mathbf{F}_{\mathrm{BB}} \, \mathbf{s}$.

Finally, $\mathbf{H}\in{\mathbb{C}^{M^2\times M^2}}$ is the channel matrix. It describes the $P$ paths provided by the environment in which the Tx and Rx arrays operate. As the signals are reflected at $P$ assumed objects, the channel matrix is modeled as a sum of weighted outer products of array propagation vectors by \cite{Sayeed_07}
\begin{equation}
\mathbf{H} = \sum_{p=1}^{P}\alpha_{p}
\big[\mathbf{a}_{\mathrm{r}}(\theta^{\mathrm{r}}_{p}, \phi^{\mathrm{r}}_{p})\big]
\big[{\mathbf{a}_{\mathrm{t}}}(\theta^{\mathrm{t}}_{p},\phi^{\mathrm{t}}_{p})\big]^{\mathrm{T}}
\cdot e^{-\frac{j 2\pi D_{p}}{\lambda}} \, .
\label{equ:phy_SISO_channel}
\end{equation}
The variables $D_{p}$ denote the path lengths between the phase centers of the transceivers for path $p$. For simplification, in later numerical evaluations, we associate each path with one reflection object, e.g., ground. In this case, the path gain along path $p$ is described as
\begin{equation}
    \mathrm{\alpha}_{p}=\Gamma_{p}\cdot \frac{\lambda}{4\pi D_{p}},
    \label{equ:pathloss}
\end{equation}
with a reflection coefficient $\Gamma_{p}$. For the LoS path, we set $\Gamma_{p}=1$. Other reflected paths are evaluated using Fresnel's formulas with the angle of incidence, dielectric constant and conductivity. Note that, we model single reflection at single object only. For reflection with a scattering cluster, more complicated models for $\mathrm{\alpha}_{p}$ can be applied \cite{Heath_2014_channel_model}. The vectors
$\mathbf{a}_{\mathrm{r}}(\theta^{\mathrm{r}}_{p},\phi^{\mathrm{r}}_{p})$ and
$\mathbf{a}_{\mathrm{t}}(\theta^{\mathrm{t}}_{p}, \phi^{\mathrm{t}}_{p})$ of size $M^2
\times 1$ are array response vectors at transmitter and receiver side, respectively. And they are parametrized by the elevation and azimuth angles of the paths that were already introduced.

Assuming the subarrays lie in the $xy$-plane, while the link distance $D$ is measured
along the $z$-coordinate, one such vector $\mathbf{a}_{\mathrm{r}}(\theta^{\mathrm{r}}_{p}, \phi^{\mathrm{r}}_{p})$ at the receiver is written as \cite{Balanis2005}
\begin{eqnarray}
\mathbf{a}_{\mathrm{r}}(\theta^{\mathrm{r}}_{p},\phi^{\mathrm{r}}_{p})\! = \big[&&\!\!\!\!\!\!\!\!\! 1,\ \dots,\ e^{ \frac{j2\pi d_\mathrm{e}}{\lambda}
(m_\mathrm{x} \sin (\theta^{\mathrm{r}}_{p}) \cos (\phi^{\mathrm{r}}_{p}) + m_\mathrm{y} \sin
(\theta^{\mathrm{r}}_{p}) \sin (\phi^{\mathrm{r}}_{p}))}, \dots, \nonumber \\ &&
\!\!\!\!\! \dots,\ e^{\frac{j2\pi d_\mathrm{e}}{\lambda} ((M-1)\sin (\theta^{\mathrm{r}}_{p}) \cos
(\phi^{\mathrm{r}}_{p}) +(M-1)\sin (\theta^{\mathrm{r}}_{p}) \sin
(\phi^{\mathrm{r}}_{p}) )}\ \big]^{\mathrm{T}},
\end{eqnarray}
where $0\leq m_\mathrm{x},\, m_\mathrm{y} < M-1$ are the $x,y$ indices of an antenna element in the subarray and
$\mathbf{a}_{\mathrm{t}}(\theta^{\mathrm{t}}_{p},\phi^{\mathrm{t}}_{p})$ can be written
in a similar fashion.

From a baseband point of view, an effective channel $\mathbf{H}_{\mathrm{eff}}$ of size $B\times B$ is seen as the physical channel including the RF frontends. Meanwhile the effective noise $\mathbf{n}_{\mathrm{eff}}$ on the RF chains are of size $B\times 1$. Writing the analog beamforming matrices $\mathbf{F}_{\mathrm{RF}}, \mathbf{W}_{\mathrm{RF}} \in \mathbb{C}^{M^2 \times B}$ as a collection of column vectors chosen from an available set of beamforming vectors ('codebook') in the RF frontends, these matrices become
\begin{equation}
\mathbf{F}_{\mathrm{RF}}=[\mathbf{f}_{1},\mathbf{f}_{2},\dots,\mathbf{f}_{B}], \quad
\mathbf{W}_{\mathrm{RF}}=[\mathbf{w}_{1},\mathbf{w}_{2},\dots,\mathbf{w}_{B}],
\end{equation}
where $\mathbf{f}_{b}, \mathbf{w}_{b}\in \mathbb{C}^{M^2\times 1}, 1\leq b \leq B$ refer
to beam pattern $b$ formed by the subarrays at transmitter and
receiver sides. It is easily seen that together with the channel describing the
environment, there occur two groups of \emph{inner} products between analog beamforming and array propagation vectors in Equ.~(\ref{equ:SISO_system}). We denote the inner products as $g^{\mathrm{t}}_{b}(\theta^{\mathrm{t}}_{p},
\phi^{\mathrm{t}}_{p})\triangleq\big[{\mathbf{a}_{\mathrm{t}}}(\theta^{\mathrm{t}}_{p},
\phi^{\mathrm{t}}_{p})\big]^{\mathrm{T}}\cdot \mathbf{f}_{b}$ and
$g^{\mathrm{r}}_{b}(\theta^{\mathrm{r}}_{p},
\phi^{\mathrm{r}}_{p})\triangleq\big[{\mathbf{a}_{\mathrm{r}}}(\theta^{\mathrm{r}}_{p},
\phi^{\mathrm{r}}_{p})\big]^{\mathrm{T}}\cdot \mathbf{w}_{b}$. The coefficient $g^{i}_{b}(\theta^{i}_{p}, \phi^{i}_{p})$ actually indicates the gain of beam pattern $b$ to radiate/collect energy over path $p$ with elevation angle $\theta^i_{p}$ and azimuthal angle $\phi^i_{p}$ at transceiver $i\in\{\mathrm{t},\ \mathrm{r}\}$ side, $1 \leq b \leq B$.

Collecting all $B$ pairs of $\{ g^{\mathrm{t}}_{b}(\theta^{\mathrm{t}}_{p},
\phi^{\mathrm{t}}_{p}),\ g^{\mathrm{r}}_{b}(\theta^{\mathrm{r}}_{p},
\phi^{\mathrm{r}}_{p}) \} $ for path $p$, two column vectors are formed to represent the array gains at transmitter and receiver side for this path. Those two gain vectors of size $B\times 1$ can be expressed as
\begin{eqnarray}
\!\!\!\!\!\!\!\!\!\!\!\!\!\!\!\!\!\! \mathbf{g}^{\mathrm{t}}(\theta^{\mathrm{t}}_{p}, \phi^{\mathrm{t}}_{p}) &\triangleq& \Big[
 \big[{\mathbf{a}_{\mathrm{t}}}(\theta^{\mathrm{t}}_{p},
\phi^{\mathrm{t}}_{p})\big]^{\mathrm{T}}\cdot \mathbf{f}_{1},\
\big[{\mathbf{a}_{\mathrm{t}}}(\theta^{\mathrm{t}}_{p},
\phi^{\mathrm{t}}_{p})\big]^{\mathrm{T}}\cdot \mathbf{f}_{2},\ \dots,\ \big[{\mathbf{a}_{\mathrm{t}}}(\theta^{\mathrm{t}}_{p},
\phi^{\mathrm{t}}_{p})\big]^{\mathrm{T}} \cdot \mathbf{f}_{B}\ \Big]^{\mathrm{T}}, \; \mathrm{and} \nonumber \\
\mathbf{g}^{\mathrm{r}}(\theta^{\mathrm{r}}_{p}, \phi^{\mathrm{r}}_{p}) &\triangleq& \Big[
 \big[{\mathbf{a}_{\mathrm{r}}}(\theta^{\mathrm{r}}_{p},
\phi^{\mathrm{r}}_{p})\big]^{\mathrm{T}}\cdot \mathbf{w}_{1},\
\big[{\mathbf{a}_{\mathrm{r}}}(\theta^{\mathrm{r}}_{p},
\phi^{\mathrm{r}}_{p})\big]^{\mathrm{T}}\cdot \mathbf{w}_{2},\ \dots,\ \big[{\mathbf{a}_{\mathrm{r}}}(\theta^{\mathrm{r}}_{p},
\phi^{\mathrm{r}}_{p})\big]^{\mathrm{T}} \cdot \mathbf{w}_{B}\ \Big]^{\mathrm{T}},
\label{equ:receiver_gains}
\end{eqnarray}
at transmitter and receiver side, respectively. Summing over all $P$ paths again leads to an effective channel $\mathbf{H}_{\mathrm{eff}}$ in baseband as
\begin{equation}
\mathbf{H}_{\mathrm{eff}}\triangleq \mathbf{W}_{\mathrm{RF}}^{\mathrm{T}} \mathbf{H} \mathbf{F}_{\mathrm{RF}} =\sum_{p=1}^{P}\alpha_{p}
\big[\mathbf{g}^{\mathrm{r}}_{}(\theta^{\mathrm{r}}_{p}, \phi^{\mathrm{r}}_{p})\big]
\big[{\mathbf{g}^{\mathrm{t}}_{}}(\theta^{\mathrm{t}}_{p},\phi^{\mathrm{t}}_{p})\big]^{\mathrm{T}}
\cdot e^{-\frac{j2\pi D_{p}}{\lambda}} \, . \label{equ:SISO_channel}
\end{equation}
Ideally $\mathbf{H}_{\mathrm{eff}}$ would be a diagonal matrix, if the analog beamformers would collect energy only from a single path while steering nulls to all other paths. In this case, one finds $g_b^\mathrm{i}(\theta^{\mathrm{i}}_{p}, \phi^{\mathrm{i}}_{p})$ satisfying $g_b^\mathrm{i}(\theta^{\mathrm{i}}_{p}, \phi^{\mathrm{i}}_{p}) = M^2\cdot \delta_{b,p}$, where $\delta_{b,p}$ indicates a Dirac impulse and becomes one if the steering direction of beam $b$ aligned with $p$-th path direction. In practice for finite main lobe width, we can only hope to suppress partially the other paths. 

Incorporating the analog beamforming with the physical channel, we arrive at the standard linear model that was originally used to describe the spatial multiplexing scenario \cite{Foschini1996}. The model can be written in terms of our variables as
\begin{equation}
\mathbf{y} = \mathbf{W}_{\mathrm{BB}}^{\mathrm{T}} \mathbf{H}_{\mathrm{eff}}
\mathbf{F}_{\mathrm{BB}}\, \mathbf{s} +\mathbf{W}_{\mathrm{BB}}^{\mathrm{T}} \mathbf{n}_{\mathrm{eff}}.
\label{equ:eff_MIMO_system}
\end{equation}
Obviously all well-known multi-user detection strategies such as linear filtering,
successive interference cancellation proposed in the literature \cite{Foschini1996} as
well as even more advanced concepts such as sphere detection approaches are applicable
as proposed in the literature. 

Please note that the effective noise $\mathbf{n}_{\mathrm{eff}}$ is not i.i.d white Gaussian noise, if the selected beam vectors $\mathbf{w}_{b}$ are non-orthogonal to each other. This is because its covariance matrix $\mathbf{R}_{\mathbf{n}_{\mathrm{eff}}}=\mathbb{E}[\mathbf{n}_{\mathrm{eff}}\mathbf{n}_{\mathrm{eff}}^{\mathrm{H}}]= \sigma_{\mathrm{n}}^2 \cdot \mathbf{W}_{\mathrm{RF}}^{\mathrm{T}} (\mathbf{W}_{\mathrm{RF}}^{\mathrm{T}})^{\mathrm{H}}$ is no longer a diagonal matrix with equal amplitudes. Only if the vectors $\mathbf{w}_{b}$ are orthogonal to each other, $\mathbf{n}_{\mathrm{eff}}$ stays i.i.d white Gaussian noise as $\mathbf{R}_{\mathbf{n}_{\mathrm{eff}}} = \sigma^2_{\mathrm{n}} M^2 \cdot \mathbf{I}_{B}$.

\section{Two-level Spatial Multiplexing with an Array of Subarrays}
\label{sec:two_level_MIMO}

\begin{figure}[t]
\centering
   \includegraphics[width=\textwidth]{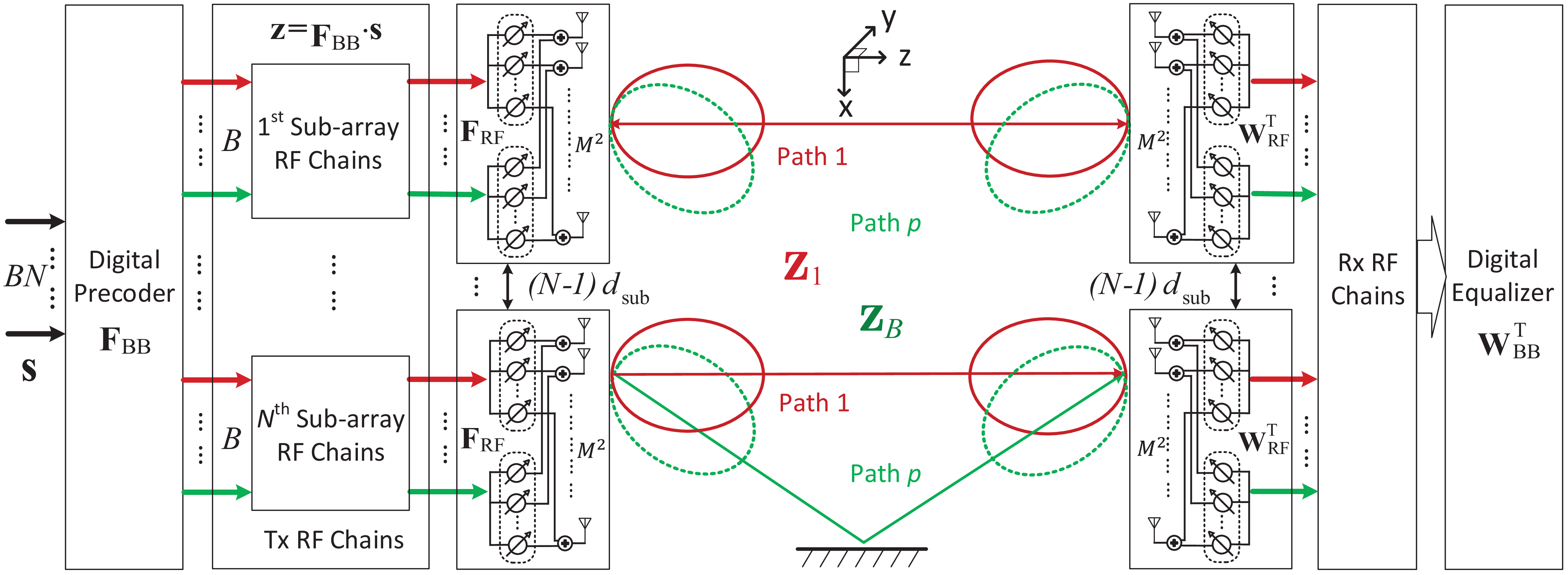}
\center \vspace{-2mm} \caption{\small Simplified block diagram of a mmWave multi-subarray MIMO system with an advanced hybrid beamforming architecture. Subarrays are large spaced, $d_\mathrm{sub} \gg \lambda$.} \label{fig:system_block_diagram_NB}
\vspace{-6mm}
\end{figure}

At this stage our proposal for a two-level spatial multiplexing concept might be already
obvious. It simply applies both transmission modes described in the previous section simultaneously and is illustrated schematically in Fig~\ref{fig:system_geometry_relations}(c). As one may recognize, hybrid beamforming (more precisely, the additional degrees of freedom due to pattern multiplication of an \emph{array} of subarrays) provides the basis to connect both approaches in a two-level hierarchical multiplexing system\footnote{While in single path (LoS) spatial multiplexing, the data streams are separated by phase differences. Additional beamforming provides separation of the streams along different paths, i.e. by angular/gain differences. Along each path, only the corresponding desired stream is dominant in magnitude, while all others produce certain weaker interference levels.}.

To keep things simple, we again only consider a standard two-path model \cite{rappaport}
that consists of a LoS path and a ground reflected path. The transceivers are
assumed to be at the same height $h$ and are separated by a link distance $D$ along the
horizontal direction. Both transmitter and receiver are assumed to be an antenna
array of several subarrays. These subarrays provide power gain on one hand and allow beam steering via their anisotropic radiation characteristics on the other hand.

When multipaths in a limited scattering environment become available, beamforming with
the help of the subarrays (for millimeter waves these require only the area in the order of $\mathrm{cm}^2$) can be employed. After applying (adaptive or training based) beam steering algorithms and addressing the directions that allow energy transfer, these reflecting paths can be excited. By putting several of the subarrays with larger distances, a similar multiplexing gain as for the LOS direction can be expected in addition to the multiplexing offered by multiple path directions. If the subarray spacing was only optimized for a single path, e.g., LoS path, it will be suboptimal for the other paths. However, due to the robustness of the scheme w.r.t. relative rotations and/or translations of the whole array, it is still expected that
some of the non-LoS (NLoS) paths support more than one spatial stream.

Let us extend the model of Section~\ref{sec:Model_single_subarray} to a two-level
spatial multiplexing system that exploits both kinds of spatial multiplexing gain, considering the block diagram in Fig.~\ref{fig:system_block_diagram_NB}. Again we assume that the transmit array and the receive array are facing each other and are arranged symmetrically over a distance $D$. As the spatial multiplexing within paths depends on the subarray arrangements, we assume that the LoS link is available and the arrays are arranged accordingly. The direct links between corresponding subarray pairs are the broadsides to the array planes. The higher level of the hierarchical MIMO system contains $N$ uniformly spaced linear arrays, so that each antenna element on the higher level is equivalent to a subarray on the lower level. All subarrays are again modeled as uniform square arrays with $M^2$ isotropically radiating elements with half wavelength spacing, such that $d_\mathrm{e}=\lambda/2$. The subarray spacing $d_{\mathrm{sub}}$ should fulfill at least approximately the condition for LoS (single path) spatial multiplexing \cite{Song_GC2015, Larsson} i.e. $d_{\mathrm{sub}} \simeq \sqrt{\lambda\cdot D/N}$ so that $\lambda \ll d_{\mathrm{sub}}\ll D$. In this work, the system is also named as multi-subarray MIMO system with large subarray separation.

To extend Equ.~(\ref{equ:SISO_system}) and (\ref{equ:SISO_channel}), we assume again
that $P$ paths are available to all subarrays. In addition, we assume that all distances
involved in the geometric relations of reflecting objects w.r.t. the transceivers
are much larger than $d_{\mathrm{sub}}$. The environment and the associated channel (coupling) between the antenna arrays can be considered as deterministic as for wireless backhaul or partially random. Furthermore, if we take into account that the ground surface will not be perfectly flat in practice, the phase relation of beam(s) for the reflected path(s) can still be acquired based on training.

Applying the same analog beamforming strategy to all $N$ subarrays, the equation we end
up with is again similar to the single subarray case as
\begin{equation}
\mathbf{y} = \mathbf{W}_{\mathrm{BB}}^{\mathrm{T}} \underbrace{\mathbf{W}_{\mathrm{RF},N}^{\mathrm{T}} \mathbf{H}\, \mathbf{F}_{\mathrm{RF},N}}_{\triangleq\mathbf{H}_{\mathrm{eff}}} \mathbf{F}_{\mathrm{BB}} \, \mathbf{s} +\mathbf{W}_{\mathrm{BB}}^{\mathrm{T}} \underbrace{\mathbf{W}_{\mathrm{RF},N}^{\mathrm{T}} \mathbf{n}}_{\mathbf{n}_{\mathrm{eff}}}, \label{equ:MIMO_system}
\end{equation}
with the only difference that now $\mathbf{s}$, $\mathbf{y}$ are $NB \times 1$ vectors of receive and transmit symbols, respectively. In this case, the number of supported data stream $N_\mathrm{s}$ is limited by the number of available paths $P$, number of RF chains at each subarray $B$ and number of subarrays $N$ with $N_\mathrm{s}\leq NP$, $N_\mathrm{s}\leq NB$.
 
Similarly, the size of $\mathbf{n} \sim \mathcal{CN}(\mathbf{0},\,\sigma^2_{n}\cdot\mathbf{I}_{N M^2})$, as well as the matrices $\mathbf{F}_{\mathrm{BB}}$, $\mathbf{W}_{\mathrm{BB}}$ that are now of size $NB \times NB$, need to be adjusted. Meanwhile, the extension of the analog beamforming matrices are denoted by $\mathbf{F}_{\mathrm{RF},N}$, $\mathbf{W}_{\mathrm{RF},N}$. As the same analog beamforming is applied at all $N$ subarrays, these matrices are related to their single subarray versions by $\mathbf{F}_{\mathrm{RF},N}= \mathbf{F}_{\mathrm{RF}}\otimes \mathbf{I}_N$ and $\mathbf{W}_{\mathrm{RF},N}= \mathbf{W}_{\mathrm{RF}} \otimes \mathbf{I}_N$ with sizes $NM^2 \times NB$ at transmitter and receiver side, respectively. Here we recall that $\otimes$ denotes the Kronecker product.

The joint analog and digital beamforming applies to transmit symbol vector $\mathbf{s}$ and actually radiated symbol vector $\mathbf{x}=\mathbf{F}_{\mathrm{RF},N} \mathbf{F}_{\mathrm{BB}}\cdot \mathbf{s}$ is now of size $NM^2 \times 1$. Meanwhile, the actually transmitted baseband symbol vector $\mathbf{z}= \mathbf{F}_{\mathrm{BB}}\cdot \mathbf{s}$ has become a block vector $\mathbf{z}=[\mathbf{z}_1^\mathrm{T}, \mathbf{z}_2^\mathrm{T},\cdots,\mathbf{z}_{B}^\mathrm{T}]^\mathrm{T}$. The $\mathbf{z}_b\in \mathbb{C}^{N\times 1}$ represents the symbols transmitted via beam $b$ that is simultaneously radiated from all subarrays.

In this section, $\mathbf{H}\in{\mathbb{C}^{NM^2\times NM^2}}$ is the channel matrix including the array response vectors to and from reflecting objects for all $N$ subarrays. For all subarrays spaced with $d_\mathrm{sub}$ under above geometry conditions, the array responses are the same\footnote{The plane wave assumption is still applicable \emph{within} the subarrays, because there the antenna elements are only separated by approximately $\lambda /2$.}, given by $\alpha_{p},\ \mathbf{a}_{\mathrm{r}}(\theta^{\mathrm{r}}_{p},\phi^{\mathrm{r}}_{p}),\ \mathbf{a}_{\mathrm{t}}(\theta^{\mathrm{t}}_{p},\phi^{\mathrm{t}}_{p})$. However, the spherical wave model needs to be applied again on the higher array level, as for LoS spatial multiplexing. The proof and further explanations on the applicable wave models of different levels can be found in the Appendix. In this way, the relative phases of the phase centers of different subarray pairs at the transceivers might be different via the propagation along the same path. Therefore, phase coupling matrix $\mathbf{H}_{p} \in \mathbb{C}^{N\times N}$ (similar to the one given for the example in Equ.~\eqref{eq:coupling_example}) should be introduced to replace the comment phase term in Equ.~\eqref{equ:phy_SISO_channel}. Combining these effects, the complete channel can be formulated as
\begin{equation}
\mathbf{H} = \sum_{p=1}^{P}\alpha_{p}
\big[\mathbf{a}_{\mathrm{r}}(\theta^{\mathrm{r}}_{p}, \phi^{\mathrm{r}}_{p})\big]
\big[{\mathbf{a}_{\mathrm{t}}}(\theta^{\mathrm{t}}_{p},\phi^{\mathrm{t}}_{p})\big]^{\mathrm{T}}
\otimes \mathbf{H}_{p}. \label{equ:phy_MIMO_channel}
\end{equation}
The elements of $\mathbf{H}_{p}$ are given again by terms of the form $\{\mathbf{H}_{p}\}_{lk}=e^{-j\frac{2\pi}{\lambda}\cdot D^{(lk)}_{p}}$, where $ D^{(lk)}_{p}$ denotes the distance between the $l$-th transmit subarray and the $k$-th receive subarray via the $p$-th path.

From a baseband point of view, an effective channel matrix $\mathbf{H}_{\mathrm{eff}}$ of size $NB \times NB$ including the analog beamformer operation in the RF frontends can be constructed again. Following the steps given in Section \ref{sec:Model_single_subarray}, two groups of inner products, between analog beam steering vectors and the array response vectors of the reflecting objects, can be formed firstly. A vector in one such group represents either transmitter or receiver side gain coefficients of all applied beam patterns at a certain path. Together with the path gain, the outer product between corresponding vectors of two different groups then represents the MIMO gain coupling matrix of a particular path. However, due to the fact that there are multiple subarrays involved, the products are conducted with commutative law of the Kronecker product $(\mathbf{A}\otimes \mathbf{B})(\mathbf{C}\otimes \mathbf{D})=(\mathbf{A}\mathbf{C}) \otimes(\mathbf{B} \mathbf{D})$ \cite{Loan2000}. After summing up all paths, an effective channel on baseband similar to Equ.~\eqref{equ:SISO_channel} can be obtained as
\begin{equation}\label{equ:effective_channel_model}
\mathbf{H}_{\mathrm{eff}}=\sum_{p=1}^{P}\alpha_{p}
\big[\mathbf{g}^{\mathrm{r}}_{}(\theta^{\mathrm{r}}_{p}, \phi^{\mathrm{r}}_{p})\big]
\big[{\mathbf{g}^{\mathrm{t}}_{}}(\theta^{\mathrm{t}}_{p},\phi^{\mathrm{t}}_{p})\big]^{\mathrm{T}}
\otimes\mathbf{H}_{p} \, .
\end{equation}
The only difference is that this effective channel does not only support $B$ streams
as for the multipath single subarray case but now we have $\mathbf{H}_{\mathrm{eff}} \in
\mathbb{C}^{NB\times NB}$, because we are using $N$ subarrays in parallel over each path
(for which the streams are discriminated by their respective phase coupling matrix).

Also note that the same argument for the effective noise $\mathbf{n}_{\mathrm{eff}}$ in Section~\ref{sec:Model_single_subarray} hold. If the vectors $\mathbf{w}_{b}$ are orthogonal to each other (i.~e. an orthogonal 'codebook' is used in the RF equalizer.) $\mathbf{n}_{\mathrm{eff}}$ is i.i.d white Gaussian noise with $\mathbf{R}_{\mathbf{n}_{\mathrm{eff}}} = \sigma^2_{\mathrm{n}} M^2 \cdot \mathbf{I}_{NB}$. However, if the selected beam vectors $\mathbf{w}_{b}$ are non-orthogonal to each other, its covariance matrix satisfies $\mathbf{R}_{\mathbf{n}_{\mathrm{eff}}}=\mathbb{E}[\mathbf{n}_{\mathrm{eff}}\mathbf{n}_{\mathrm{eff}}^{\mathrm{H}}]= \sigma_{\mathrm{n}}^2 \cdot [(\mathbf{W}_{\mathrm{RF}}^{\mathrm{T}} (\mathbf{W}_{\mathrm{RF}}^{\mathrm{T}})^{\mathrm{H}})\otimes \mathbf{I}_{\mathrm{N}}]$.

\subsection{Example: Two-level spatial multiplexing over two paths using two subarrays}

In Sec.~\ref{sec:Results_and_Discussion}, our numerical evaluations will be carried out
for a system with $N=2$ subarrays communicating over a standard two-path model
\cite{rappaport}, including a LoS path and an additional ground reflected path (see
Fig.~\ref{fig:system_geometry_relations}(c)). Therefore, we subsequently illustrate, how
the effective channel $\mathbf{H}_{\mathrm{eff}}$ is obtained for this example case. For simplification, we use $B=P=2$ beams at each subarray which can excite the two paths with equal number of beams in the ideal cases.

Let the transceivers be at the same height $h$. It is also further assumed that $\lambda \ll d_\mathrm{sub} \ll h < D$, as for MIMO systems with large antenna separation. On the lower level, the analog beamforming algorithm at all subarrays orients two beams, one excites the LoS path and the other targets at the ground reflected path. On the higher level, the phase relations of the coupling matrices are determined by the lengths of the propagation paths $D^{(lk)}_{p}$. Meanwhile, the lengths are determined by the geometry of the paths as well as the arrangements of the antenna arrays at transmitter and receiver sides. Furthermore, with the assumption of $\lambda \ll d_\mathrm{sub}\ll h < D$, the angle differences for different antenna subarrays are smaller than $\arctan [(N-1)\cdot d_{\mathrm{sub}}/D] \simeq \sqrt{\lambda/D} \approx 0$. Therefore, we assume that the array gains, which are observed by the different subarrays via the same path (LoS or ground reflection) and the same beam pattern, are equal. However, for the same chosen beam pattern, the gains read differently along different path directions as $h$ is of the same order as $D$. This is because the angle difference is of the order $\arctan [h/D]$ which is no longer negligible as shown in Fig.~\ref{fig:system_geometry_relations}.

Using the above assumptions with $B=2$, $P=2$, $N=2$, the effective channel including the RF frontends can be written as
\begin{eqnarray}\label{equ:effective_channel_model_two_path}
\!\!\!\!\! &&\!\!\!\!\!\!\! \mathbf{H}_{\mathrm{eff}}= \alpha_{1} \cdot \left [\! \begin{array}{cc}
g^{\mathrm{t}}_1(\theta^{\mathrm{t}}_{1},\phi^{\mathrm{t}}_{1})g^{\mathrm{r}}_1(\theta^{\mathrm{r}}_{1},
\phi^{\mathrm{r}}_{1})&g^{\mathrm{t}}_2(\theta^{\mathrm{t}}_{1},\phi^{\mathrm{t}}_{1})
g^{\mathrm{r}}_1(\theta^{\mathrm{r}}_{1},\phi^{\mathrm{r}}_{1})\\g^{\mathrm{t}}_1(\theta^{\mathrm{t}}_{1},
\phi^{\mathrm{t}}_{1})g^{\mathrm{r}}_2(\theta^{\mathrm{r}}_{1},\phi^{\mathrm{r}}_{1})&
g^{\mathrm{t}}_2(\theta^{\mathrm{t}}_{1},\phi^{\mathrm{t}}_{1})g^{\mathrm{r}}_2(\theta^{\mathrm{r}}_{1},
\phi^{\mathrm{r}}_{1})\end{array} \!\right] \!\otimes\!\underbrace{\left [\! \begin{array}{cc}
1& e^{-j\pi/2} \\ e^{-j\pi/2}& 1 \end{array} \!\right] \cdot e^{-j2\pi D/\lambda} }_{\mathbf{H}_{1}:\ \mathbf{H}_{1}=\mathbf{H}_{\mathrm{LoS}}}  +\, \alpha_{2}\cdot  \nonumber \\
 \!\!\!\!\! && \!\!\!\!\!\!\!  \left [\! \begin{array}{cc}g^{\mathrm{t}}_1(\theta^{\mathrm{t}}_{2},
\phi^{\mathrm{t}}_{2})g^{\mathrm{r}}_1(\theta^{\mathrm{r}}_{2},\phi^{\mathrm{r}}_{2})&
g^{\mathrm{t}}_2(\theta^{\mathrm{t}}_{2},\phi^{\mathrm{t}}_{2})g^{\mathrm{r}}_1(\theta^{\mathrm{r}}_{2},
\phi^{\mathrm{r}}_{2})\\g^{\mathrm{t}}_1(\theta^{\mathrm{t}}_{2},\phi^{\mathrm{t}}_{2})
g^{\mathrm{r}}_2(\theta^{\mathrm{r}}_{2},\phi^{\mathrm{r}}_{2})&g^{\mathrm{t}}_2(\theta^{\mathrm{t}}_{2},
\phi^{\mathrm{t}}_{2})g^{\mathrm{r}}_2(\theta^{\mathrm{r}}_{2},\phi^{\mathrm{r}}_{2})\end{array}\! \right]\!\otimes
\!\underbrace{\left [\! \begin{array}{cc}
e^{-j \frac{2\pi\sqrt{(2h+d_{\mathrm{sub}})^2+D^2}}{\lambda} } & \!\!\! e^{-j\frac{2\pi}{\lambda} \sqrt{(2h)^2+D^2}} \\ e^{-j\frac{2\pi}{\lambda}\sqrt{(2h)^2+D^2}}& \!\!\! e^{-j \frac{2\pi\sqrt{(2h-d_{\mathrm{sub}})^2+D^2}}{\lambda} }  \end{array} \!\right]}_{\mathbf{H}_{2}:\ \{\mathbf{H}_{2}\}_{lk}=e^{-j 2\pi D^{(lk)}_2/\lambda}}.
\end{eqnarray}
Note that Equ.~(\ref{equ:effective_channel_model_two_path}) is a special case of Equ.~(\ref{equ:effective_channel_model}) with $\mathbf{H}_1, \, \mathbf{H}_2$ of size $N \times N$, $N=2$. The gain matrices that occur as the
first factors in the Kronecker products are of size $B \times B$, $B=2$. Their entries are obtain via the outer product of two vectors $\mathbf{g}^{\mathrm{r}}(\theta^{\mathrm{r}}_{p},\phi^{r}_{p})$, $\mathbf{g}^{\mathrm{t}}(\theta^{\mathrm{t}}_{p},\phi^{t}_{p})$. Each vector is obtained according to Equ.~(\ref{equ:receiver_gains}) as
\begin{equation}
\mathbf{g}^{i}(\theta^{i}_{p},\phi^{i}_{p})=\left [\begin{array}{cc}
g^{i}_1(\theta^{i}_{p},\phi^{i}_{p})\\ g^{i}_2(\theta^{i}_{p},\phi^{i}_{p})
\end{array}\right] \,,
\end{equation}
with $i \in \{ \mathrm{t}, \mathrm{r}\}$, so that $\mathbf{H}_\mathrm{eff}$ is of size $4 \times 4$.

\textbf{Array patterns of the subarrays}:
To fully specify $\mathbf{H}_\mathrm{eff}$, we work out exemplary radiation patterns for
USAs consisting of $M^2$ antenna elements with element spacing $d_\mathrm{e}=\lambda/2$ and isotropically radiating elements. For the RF precoder $\mathbf{F}_{\mathrm{RF}}$ and the RF equalizer $\mathbf{W}_{\mathrm{RF}}$, an implementation using analog phase shifters (see e.g., Ref.~\cite{rf}) is assumed. These provide different progressive phase shifts among the
antenna signals for different steering angles. With the phase increments given by
$\beta^{\mathrm{t}}_{x,b}$ ($\beta^{\mathrm{t}}_{y,b}$) and $\beta^{\mathrm{r}}_{x,b}$
($\beta^{\mathrm{r}}_{y,b}$) between adjacent antenna signals along $x$- and
$y$-directions represented by the columns $\mathbf{f}_{b}$ and $\mathbf{w}_{b}$,
respectively, we get
\begin{equation}
\mathbf{f}_{b} = \big[1,\ \dots,\ e^{j (m_\mathrm{x}\beta^{\mathrm{t}}_{x,b} + m_\mathrm{y} \beta^{\mathrm{t}}_{y,b})},\ \dots,\ e^{j((M-1) \beta^{\mathrm{t}}_{x,b}+(M-1) \beta^{\mathrm{t}}_{y,b} )} \big]^{\mathrm{T}},
\end{equation}
and
\begin{equation}
\mathbf{w}_{b} = \big[1,\ \dots,\ e^{j (m_\mathrm{x}\beta^{\mathrm{r}}_{x,b} + m_\mathrm{y} \beta^{\mathrm{r}}_{y,b})},\ \dots,\ e^{j((M-1) \beta^{\mathrm{r}}_{x,b}+(M-1) \beta^{\mathrm{r}}_{y,b} )} \big]^{\mathrm{T}} \, .
\end{equation}
Using Equ.~(\ref{equ:receiver_gains}), the gains of the
transmit/receive subarrays using beam $b$ are expressed as
\begin{eqnarray}\label{b}
g^{i}_b(\theta^{i}_{p},\phi^{i}_{p}) &=& \left \{ \begin{array}{ccc}
\big[{\mathbf{a}_{\mathrm{t}}}
(\theta^{\mathrm{t}}_{p},\phi^{\mathrm{t}}_{p})\big]^{\mathrm{T}}\cdot &\!\!\!\!\!
\mathbf{f}_{b}\ , & \text{if } i=\mathrm{t}; \\
\big[{\mathbf{a}_{\mathrm{r}}}(\theta^{\mathrm{r}}_{p},\phi^{\mathrm{r}}_{p})\big]^{\mathrm{T}}\cdot
&\!\!\!\! \mathbf{w}_{b}, & \text{if } i=\mathrm{r}; \end{array} \right . \nonumber \\
&=& \frac{\sin( \frac{M}{2}
\psi^i_{\mathrm{x}})}{\sin(\frac{\psi_{\mathrm{x}}}{2})} \frac{\sin( \frac{M}{2}
\psi^i_{\mathrm{y}})}{\sin(\frac{\psi_{\mathrm{y}}}{2})} ,
\end{eqnarray}
where
\begin{eqnarray}
\psi^i_{\mathrm{x}}&=&\frac{2\pi}{\lambda} d_{\mathrm{e}}\sin\theta^{i}_{p}
\cos\phi^i_{p}+\beta^i_{x,b}, \label{eq:angle_difference_x} \\
\psi^i_{\mathrm{y}}&=&\frac{2\pi}{\lambda} d_{\mathrm{e}}\sin\theta^{i}_{p}
\sin\phi^i_{p}+\beta^i_{y,b},
\label{eq:angle_difference_y}
\end{eqnarray}
correspond to deviations between steering angle of beam $b$, and the angles of departure/arrival of signals over path $p$. The whole derivation leading to Equ.~\eqref{eq:angle_difference_x} and Equ.~\eqref{eq:angle_difference_y} also follows the steps and results given in \cite{Balanis2005}, as $g^{i}_{b}(\theta^{i}_{p},\phi^{i}_{p})$ indicates the gain of the $b$-th beam pattern obtained for elevation angle $\theta^i_{p}$ and azimuth angle $\phi^i_{p}$ at transceiver side $i\in\{\mathrm{t},\ \mathrm{r}\}$.

\begin{figure}[t]
\centering
\includegraphics[width=0.50\textwidth]{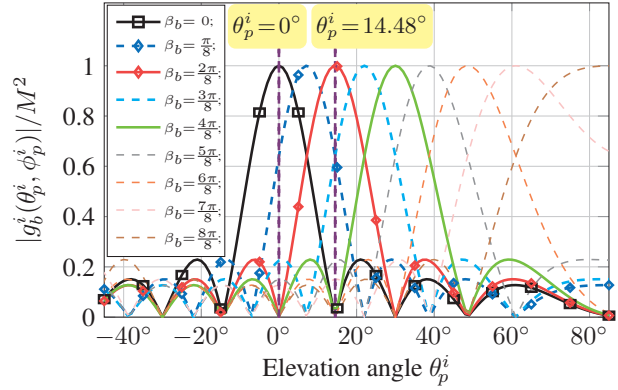}
\center
\vspace*{-3mm}
\caption{\small Normalized array patterns on the elevation plane ($\phi^i_{p}=0$) for a square subarray with $M^2=64$ antennas and a codebook of size 16 for the elevation directions.}
\label{arraypattern}
\vspace*{-3mm}
\end{figure}

Fig. \ref{arraypattern} illustrates the situation with normalized patterns in the elevation plane of a $8\times 8$-USA for which the same patterns occur in the orthogonal plane. The phased array system we assume in later evaluation contains $16$ 'codewords' ($\triangleq$ candidate beams) for analog beamforming on the elevation plane. The codewords are obtained with different phase increments $\beta_b$ (Specifically, we use multiples of $\pi/8$ in the range $(-\pi, \pi]$ associated with different steering angles equal to $\arcsin(n/8), n \in [-7, -6, \ldots 8]$).

As the later evaluations will be carried out for a wireless backhaul system with a LoS path and an additional ground reflected path, see Fig.~\ref{fig:system_geometry_relations}(c), we are only concerning beam steering in the elevation plane. Therefore, the antenna elements perpendicular to the elevation plane have no phase differences, i.e. $\beta^i_{y,b}=0$, and no beam steering needs to be applied.
Furthermore, we can take advantage of the mirror symmetry of the situation which allows to set the steering angles at transmitter and receiver to the same values. To steer a beam towards array normal $\theta^i_{p}=0$ (corresponding to the LoS path for arrays facing each other), we can simply choose our first beam pattern using $\beta^i_{x,b}=0$. Therefore, this beam pattern, denoted as $\beta_{1}=0$ for short, is always used as the first beam for exciting the LOS path. Depending on the particular values of $h$, the first pattern couples certain energy into the ground reflected path as well. To exploit the potential spatial multiplexing offered by the second path, the second beam pattern with another steering angle $\beta^i_{x,b} \neq 0$ is used. This value is then a variable and denoted as $\beta_{2}$ later for short.

\section{Spectral Efficiency with Total Transmit Power Constraint}\label{sec:spectral_Efficiency_Evaluation}

In this section, we are seeking the baseband precoders/equalizers that maximize the spectral efficiency when RF precoding/equalizing is used under a sum power constraint. This maximized spectral efficiency is an intermediate step which acts as the benchmark for capacity evaluation. The final target of this work is to show the capacity improvements by combining the two kinds of spatial multiplexing.

Our spectral efficiency evaluation is carried out assuming that Gaussian symbols are transmitted. Given by \cite{Heath_2014_channel_model}, the spectral efficiency for a joint RF/Baseband design is given by
\begin{equation}\label{rate_full}
R = \mathrm{log}_{2}[\mathrm{det}(\textbf{I}_{NB} +\mathbf{R}_{\mathrm{n}}^{-1} (\mathbf{W}_{\mathrm{BB}}^{\mathrm{T}} \mathbf{W}_{\mathrm{RF},N}^{\mathrm{T}} \mathbf{H}\, \mathbf{F}_{\mathrm{RF},N} \mathbf{F}_{\mathrm{BB}}) \mathbf{R}_{\mathbf{s}} (\mathbf{W}_{\mathrm{BB}}^{\mathrm{T}} \mathbf{W}_{\mathrm{RF},N}^{\mathrm{T}} \mathbf{H}\, \mathbf{F}_{\mathrm{RF},N} \mathbf{F}_{\mathrm{BB}})^{\mathrm{H}})],
\end{equation}
where $\mathbf{R}_{\mathrm{n}}=\sigma^2_{\mathrm{n}} (\mathbf{W}_{\mathrm{BB}}^{\mathrm{T}} \mathbf{W}_{\mathrm{RF},N}^{\mathrm{T}}) (\mathbf{W}_{\mathrm{BB}}^{\mathrm{T}} \mathbf{W}_{\mathrm{RF},N}^{\mathrm{T}})^{\mathrm{H}}$ and $\mathbf{R}_{\mathbf{s}}= \mathbb{E}[\mathbf{s} \mathbf{s}^{\mathrm{H}}]$. Meanwhile, the radiated power satisfies
\begin{equation}\label{power_full}
 \mathbb{E}[\mathbf{x}^{\mathrm{H}} \mathbf{x} ]= \mathrm{tr} (\mathbb{E}[\mathbf{x} \mathbf{x}^{\mathrm{H}}])= \mathrm{tr} (\mathbf{F}_{\mathrm{RF},N} \mathbf{F}_{\mathrm{BB}} \mathbf{R}_{\mathbf{s}} \mathbf{F}_{\mathrm{BB}}^{\mathrm{H}}\mathbf{F}_{\mathrm{RF},N}^{\mathrm{H}})\leq P_\mathrm{C},
\end{equation}
where $P_\mathrm{C}$ indicates the power constraint. To simplify the discussion later, we assume that the transmitted symbols ($\mathbf{s}$) are i.i.d variables which make $\mathbf{R}_{\mathbf{s}}$ a diagonal matrix. Without loss of generality, we assume $\mathbf{R}_{\mathbf{s}}=\mathbf{I}_{NB}$.

Maximizing the spectral efficiency requires a joint optimization over the matrices $\{\mathbf{W}_{\mathrm{BB}}^{\mathrm{T}}$, $\mathbf{W}_{\mathrm{RF},N}^{\mathrm{T}}$, $\mathbf{F}_{\mathrm{RF},N}$, $\mathbf{F}_{\mathrm{BB}}\}$. Under a total power constraint and considering the RF precoder/equalizer are taken from quantized codebooks $\{\mathscr{F}_{\mathrm{RF}}$, $\mathscr{W}_{\mathrm{RF}}\}$, the optimization can be formulated in an outer-inner problem form as \cite{Heath_2016_waterfilling}
\begin{equation}\label{equ:capacity_real}
\tilde{C} = \max_{\mathbf{F}_{\mathrm{RF}}\in \mathscr{F}_{\mathrm{RF}},\ \mathbf{W}_{\mathrm{RF}}\in \mathscr{W}_{\mathrm{RF}}}
 \left\{\! \begin{array}{c}  \max\limits_{\mathbf{F}_{\mathrm{BB}},\ \mathbf{W}_{\mathrm{BB}}} \ \ \ \ \ R   \\
\mathrm{s.t.}\ \  ||\mathbf{F}_{\mathrm{RF},N} \mathbf{F}_{\mathrm{BB}}||^2_{F} \leq P_\mathrm{C} \end{array} \right\},
\end{equation}
where the outer maximization is chosen over finite codebooks and the RF precoder/equalizer are assumed to be of $B$ RF chains at each subarray. The inner maximization is applied known the $\mathbf{F}_{\mathrm{RF}}$ and $\mathbf{W}_{\mathrm{RF}}$. Here we recall that $\mathbf{F}_{\mathrm{RF},N}= \mathbf{F}_{\mathrm{RF}}\otimes \mathbf{I}_N$ and $\mathbf{W}_{\mathrm{RF},N}= \mathbf{W}_{\mathrm{RF}} \otimes \mathbf{I}_N$.

If the RF precoder/equalizer includes non-orthogonal beam patterns, the inner maximization can not be given by standard singular value decomposition (SVD) based waterfilling algorithm on the effective channel with RF precoder/equalizer, $\mathbf{W}_{\mathrm{RF},N}^{\mathrm{T}} \mathbf{H}\, \mathbf{F}_{\mathrm{RF},N}$\footnote{For orthogonal RF precoders, SVD based waterfilling algorithm can be applied, as the radiated symbol vector $\mathbf{x}\triangleq\mathbf{F}_{\mathrm{RF},N} \mathbf{F}_{\mathrm{BB}} \, \mathbf{s}$ and the baseband noise $\mathbf{n}_{\mathrm{eff}}$ stay uncorrelated. However, the power constraint of the baseband precoder should be $M^2$ times smaller than that of the RF front-ends, as the baseband signals are radiated $M^2$ times by RF precoders.}. This is because the coupling between the baseband and analog processing must be considered. By applying SVD on the effective channel, a set of parallel subchannels appears in the transformed space. When non-orthogonal patterns are used, the RF precoder scales the located power on different subchannel differently and makes the sum power constraint to be a weighted sum power constraint. Resulting from the non-orthogonal patterns at the RF equalizer, another effect is that uncorrelated noise becomes correlated.

We define the achievable spectrum efficiency of inner maximization as $C$. Given $\mathbf{W}_{\mathrm{RF},N}^{\mathrm{T}}$ and $\mathbf{F}_{\mathrm{RF},N}$, the maximization problem becomes
\begin{eqnarray}\label{equ:capacity_full}
C &=& \max_{\mathbf{F}_{\mathrm{BB}},\ \mathbf{W}_{\mathrm{BB}}}\ \ \mathrm{log}_{2}[\mathrm{det}(\mathbf{I}_{NB} +\mathbf{F}_{\mathrm{BB}}^{\mathrm{H}}\mathbf{H}_{\mathrm{eff}}^{\mathrm{H}} (\mathbf{W}_{\mathrm{BB}}^{\mathrm{T}})^{\mathrm{H}} \mathbf{R}_{\mathrm{n}}^{-1} \mathbf{W}_{\mathrm{BB}}^{\mathrm{T}} \mathbf{H}_{\mathrm{eff}} \mathbf{F}_{\mathrm{BB}} )], \\
&&\ \ \ \ \ \ \ \ \ \ \ \ \mathrm{s.t.}\ \ ||\mathbf{F}_{\mathrm{RF},N} \mathbf{F}_{\mathrm{BB}}||^2_{F} \leq P_\mathrm{C}. \nonumber
\end{eqnarray}
Note that, for deterministic channels like a wireless backhaul channel, finding optimal $\mathbf{W}_{\mathrm{RF},N}^{\mathrm{T}}$, $\mathbf{F}_{\mathrm{RF},N}$ over a finite set is not a crucial issue. The training period using algorithms like exhaustive search is no longer limiting the system performance. Additionally, in later evaluations, the directions of the available paths are assumed to be known or approximate known to the system. Then, we assume that each path is associated with one selected beam $B=P$. Ideally, the selected beams should include the direction of the corresponding path within their main lobes. With the above assumptions, we can simplify this two-step optimization problem to the inner maximization problem, as we escape the step finding the best $\mathbf{W}_{\mathrm{RF},N}^{\mathrm{T}}$, $\mathbf{F}_{\mathrm{RF},N}$.

The baseband precoder/equalizer pair that solves the above inner maximization problem is given by the work \cite{Heath_2016_waterfilling} and we extend it for a multi-subarray scenario as proposed earlier. Both issues, the modified power constraint for hybrid beamforming in contrast to digital beamforming as well as the correlated noise, are solved by introducing one additional step before equivalent baseband precoder/equalizer. By removing the correlation of power/noise, standard SVD based waterfilling algorithm can be applied for the equivalent baseband precoder/equalizer. 

The baseband precoder $\mathbf{F}_{\mathrm{BB}}$ and equalizer $\mathbf{W}_{\mathrm{BB}}^{\mathrm{T}}$ that are capable of solving the optimization problem in Equ.~(\ref{equ:capacity_full}) are given as
\begin{equation}\label{equ:baseband_precoder}
\mathbf{F}_{\mathrm{BB}}=(\mathbf{F}_{\mathrm{RF},N}^{\mathrm{H}}\mathbf{F}_{\mathrm{RF},N})^{-\frac{1}{2}} \mathbf{V} \mathbf{\Psi},
\end{equation}
\begin{equation}\label{equ:baseband_equalizer}
\mathbf{W}_{\mathrm{BB}}^{\mathrm{T}}=\mathbf{U}^{\mathrm{H}}[\mathbf{W}_{\mathrm{RF},N}^{\mathrm{T}}(\mathbf{W}_{\mathrm{RF},N}^{\mathrm{T}})^{\mathrm{H}}]^{-\frac{1}{2}},
\end{equation}
where the diagonal matrix $\mathbf{\Psi}=\mathrm{diag}\{\psi_{1}, \psi_{2}, \ldots, \psi_{NB}\}$ contains gain coefficients that affect the power allocation. The matrices $\mathbf{U}$, $\mathbf{V}$ are unitary matrices of size $NB \times NB$ and are obtained by an SVD on the extended channel $\{[\mathbf{W}_{\mathrm{RF},N}^{\mathrm{T}}(\mathbf{W}_{\mathrm{RF},N}^{\mathrm{T}})^{\mathrm{H}}]^{-\frac{1}{2}} \mathbf{H}_{\mathrm{eff}} (\mathbf{F}_{\mathrm{RF},N}^{\mathrm{H}}\mathbf{F}_{\mathrm{RF},N})^{-\frac{1}{2}}\}$ as
\begin{equation}\label{evd}
[\mathbf{W}_{\mathrm{RF},N}^{\mathrm{T}}(\mathbf{W}_{\mathrm{RF},N}^{\mathrm{T}})^{\mathrm{H}}]^{-\frac{1}{2}} \mathbf{H}_{\mathrm{eff}} (\mathbf{F}_{\mathrm{RF},N}^{\mathrm{H}}\mathbf{F}_{\mathrm{RF},N})^{-\frac{1}{2}} =\mathbf{U}\mathbf{\Sigma} \mathbf{V}^{\mathrm{H}},
\end{equation}
where $\mathbf{\Sigma}\in{\mathbb{C}^{NB \times NB}}$ is a diagonal matrix with singular values
\begin{equation}\label{eigenvalues}
\sigma_1 \geq \sigma_2 \geq \ldots \geq \sigma_{NB} \geq 0.
\end{equation}

By using the above baseband precoder and equalizer\footnote{In this case, the received signal becomes $\mathbf{y} = \mathbf{\Sigma}\mathbf{\Psi}\mathbf{s}+\mathbf{\tilde{n}}$, with $\mathbf{\tilde{n}}= \mathbf{U}^{\mathrm{H}}[\mathbf{W}_{\mathrm{RF},N}^{\mathrm{T}}(\mathbf{W}_{\mathrm{RF},N}^{\mathrm{T}})^{\mathrm{H}}]^{-\frac{1}{2}} \mathbf{W}_{\mathrm{RF},N}^{\mathrm{T}} \mathbf{n} \sim \mathcal{CN}(\mathbf{0}, \sigma^2_{\mathrm{n}} \cdot\mathbf{I}_{NB})$ denotes the received noise.}, the optimization problem in Equ.~(\ref{equ:capacity_full}) becomes
\begin{eqnarray}\label{equ:capacity_waterfilling_matrix}
C &=& \max_{\mathbf{\Psi}}\ \ \mathrm{log}_{2}[\mathrm{det}(\mathbf{I}_{NB} +\frac{1}{\sigma_{\mathrm{n}}^2}\mathbf{\Sigma}^{\mathrm{2}}\mathbf{\Psi}^{\mathrm{2}})], \\
&&\ \ \ \ \ \ \ \mathrm{s.t.}\ \ ||\mathbf{\Psi}||^2_{F} \leq P_\mathrm{C}. \nonumber
\end{eqnarray}
Let us define a matrix $\mathbf{P}\triangleq \mathbf{\Psi}^{2}$. $\mathbf{P}$'s $q^{\mathrm{th}}$ diagonal entry $P_{q}$ represents the $q^{\mathrm{th}}$ diagonal value of the matrix $\mathbf{P}$. Therefore, the optimization problem in Equ.~(\ref{equ:capacity_waterfilling_matrix}) becomes
\begin{eqnarray}\label{capacity_waterfilling}
C &=& \max_{\mathbf{P}}\ \ \sum_{q=1}^{NB} \mathrm{log}_{2}(1 +\frac{\sigma_q^2}{\sigma_{\mathrm{n}}^2} P_{q}), \\
&&\mathrm{s.t.}\ \ \sum_{q=1}^{NB} P_{q}\leq P_\mathrm{C},\ P_{q}\geq 0. \nonumber
\end{eqnarray}

The solution for the value of $P_{q}$ is given by the waterfilling algorithm \cite{CoverThomas} as
\begin{equation}\label{Power_waterfilling}
P_{q}= \left[\kappa-\frac{\sigma_{\mathrm{n}}^2}{\sigma_{q}^2}\right]^{+},
\end{equation}
where $\kappa$ is the 'water level' and is chosen such that $\sum_q P_{q} = P_\mathrm{C}$. The notation $[x]^+$ is used for taking non-negative values only as $\max(x, 0)$. Consequently, if $\kappa-\sigma_{\mathrm{n}}^2/\sigma_{q}^2<0$, we set $P_{q}=0$. Meanwhile, in the evaluation later, the SNR $\gamma_q$ on the $q$-th subchannel is defined as
\begin{equation}\label{SNR}
\gamma_q \triangleq  \sigma_{q}^2 P_{q} /\sigma_{\mathrm{n}}^2.
\end{equation}

\section{Numerical Results for a Deterministic 2-Path Scenario}
\label{sec:Results_and_Discussion}

We evaluated the spectral efficiency achieved by a two-level spatial multiplexing system as described above. It is a deterministic 2-path channel for which LoS and ground reflected paths occur. The system involves the hybrid beamforming architecture for mmWave communication as described in Section \ref{sec:two_level_MIMO}. The subarrays are sufficiently spaced apart. This forms a LoS MIMO system with two subarrays that can take advantage of the 2-nd path in addition to the LoS path ($\{N,P\}=\{2,2\}$). For comparison, the performance of a single path single subarray system (AWGN channel with $\{N,P\}=\{1,1\}$), a two-path \emph{single} subarray system ($\{N,P\}=\{1,2\}$), and a single-path LoS MIMO system with two subarrays ($\{N,P\}=\{2,1\}$) are evaluated under the same constraints.

\textbf{Environment parameters:} Evaluations are carried out for a single subarray system ($N=1$) and a symmetric system with two uniformly spaced linear subarrays ($N=2$) aligned along the vertical direction, as shown in Fig.~\ref{fig:system_geometry_relations}. The transceivers are assumed to be separated by a transmit distance $D=100\ \mathrm{m}$ (e.g., wireless backhaul) and to be at the same height, $h\in [5,\ 35]\ \mathrm{m}$. One reflected path from the ground is assumed and the point of reflection is in the middle between the transceivers. Furthermore, the coefficient $\Gamma_2$ follows the Fresnel reflection factor with the perpendicular polarization or TE incidence \cite{rappaport}. The relative dielectric constant $\epsilon_\mathrm{r}=3.6478$ and loss tangent $\tan \delta=0.2053$ of concrete \cite{concrete} are chosen to represent ground. The coefficient $\Gamma_1$ for the LoS path is of value $\Gamma_1=1$.

\textbf{System parameters:} The subarrays are assumed to be $8\times 8$ $\lambda/2$-spaced square arrays with isotropic elements (highest antenna gain 18~dBi). The system uses a carrier frequency of 60~GHz ($\lambda=5\ \mathrm{mm}$). For transceivers with 2 subarrays, this leads to an inter subarray distance $d_{\mathrm{sub}}=\sqrt{\lambda D/N}=0.5\ \mathrm{m}$ for optimizing the spectral efficiency over the direct path\footnote{The overall spectral efficiency of system can be further increased via antenna topology optimization using the concept in \cite{Song_2015} as the arrangement must be optimized jointly for multiple directions.}. The system setup approaches the required assumption $\lambda \ll d\ll h<D$. To simplify later discussion, let us assume that the analog beamforming algorithm of subarrays orients one beam per available path, $B=P$. The codebook, from which the steering elevation angles of the subarrays could be selected, was assumed to be of size 16. $\beta_1\!=\!0$ is used for the LoS path and the positive progressive angles are used for exciting the reflected path $\beta_2 \in \{\frac{\pi}{8},\frac{2\pi}{8},\frac{3\pi}{8},\frac{4\pi}{8}\}$ in an antenna height range $h\in [5,\ 35]$ meters. The allowed bandwidth regulated by \cite{IEEE-Std-802.11ad} is of value $W=2.16~\mathrm{GHz}$. Meanwhile, the transmit power $P_\mathrm{T}$ varies from $5~\mathrm{dBm}$ to $25~\mathrm{dBm}$ in later evaluations. Considering $K$ subcarriers\footnote{$K$ is large enough that the subcarrier bandwidth is smaller than the coherent bandwidth. The spectrum efficiency is independent of $K$.}, the noise power for one subcarrier is assumed to be $\sigma_\mathrm{n}^2=k_\mathrm{B}TFW/K$, where $k_\mathrm{B}$ is the Boltzmann constant, $T=300\ \mathrm{K}$ is the absolute temperature in Kelvin, and $F=5\ \mathrm{dB}$ is the noise figure. For each subcarrier, the power constraint $P_\mathrm{C}$ is then calculated as $P_\mathrm{C}=P_\mathrm{T}/K$.

\textbf{A Link Budget:} A brief link budget is made here to offer a better understanding of our parameter settings. The allowed peak Equivalent Isotropically Radiated Power (EIRP) \cite{FCCEIRP} is of value 43~dBm at antenna gain of $18$~dBi. Considering the further transmit power degradation due to poor peak-to-average power ratio, we evaluate the average transmit power $P_\mathrm{T}$ in a range from $5~\mathrm{dBm}$ to $25~\mathrm{dBm}$. The noise power for the complete bandwidth $W$ is found as $-75.5$~dBm. The free space pathloss of LOS path according to Friis transmission equation $(\frac{\lambda}{4\pi D})^2$ is $-108$dB. Considering two parallel AWGN channels derived from a LOS MIMO system with two subarrays and $P_\mathrm{T}=20$~dBm, the calculated SNR is about 23~dB and the corresponding spectrum efficiency is 15.7~bits/s/Hz. Regarding the reflected path, the free space pathloss of the reflected path has additional loss up-to 2dB and power loss of reflection changes from 1~dB to 5~dB in the range of $h$ under investigation.

\subsection{Performance of 2-Path Spatial Multiplexing using a Single Subarray} \label{SISO_result}

\begin{figure}
\centering
\begin{minipage}{.47\textwidth}
\centering
\includegraphics[width=\linewidth]{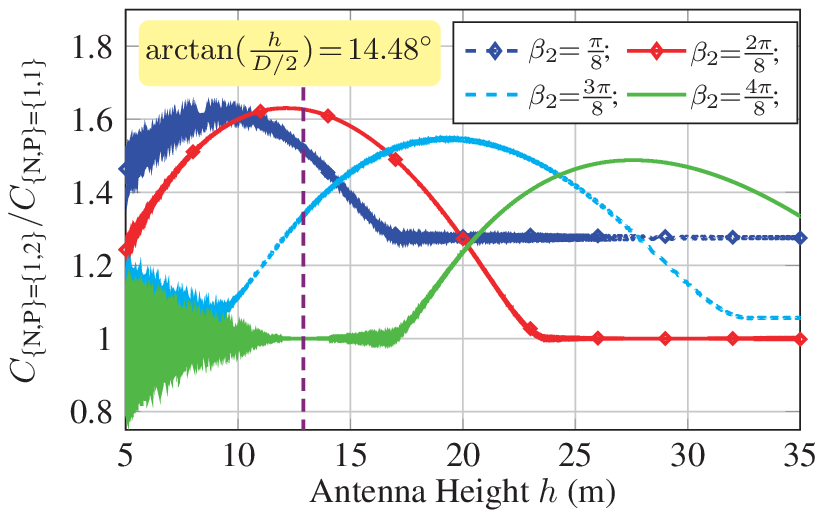}
\center
\vspace{-2mm}
\caption{\small Normalized spectral efficiencies of a 2-path single subarray system w.r.t the capacity of a single path system ($N=1$, $M=8$, $P_\mathrm{T}= 20 \mathrm{dBm}$).}
\label{singreflSISOP20}
\vspace{-3mm}
\end{minipage}
\hspace{3mm}
\begin{minipage}{.47\textwidth}
\centering
\includegraphics[width=\linewidth]{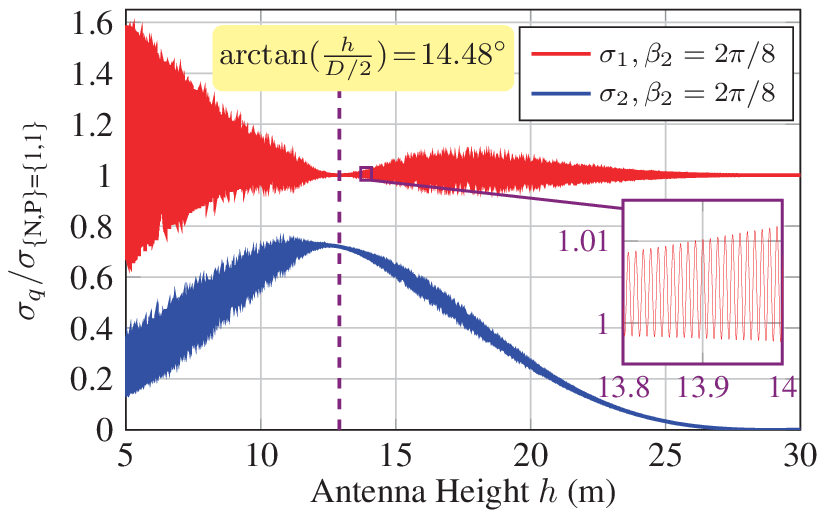}
\center
\vspace{-2mm}
\caption{\small Normalized singular values of a 2-path single subarray system w.r.t the channel gain of a single path system ($N=1$, $M=8$).}
\label{singlosSISO}
\vspace{-3mm}
\end{minipage}
\end{figure}

Fig.~\ref{singreflSISOP20} compares the relative spectral efficiencies $C_{\scalebox{.7}{\{N,P\}=\{1,2\}}}$ of a 2-path channel when different analog beam patterns are used. Meanwhile, in order to present the singular value variation of the channel, the beam pair $\{\beta_1, \beta_2\}=\{0, 2\pi/8\}$ is selected as an example in Fig.~\ref{singlosSISO}. All the values are normalized w.r.t. spectral efficiency $C_{\scalebox{.7}{\{N,P\}=\{1,1\}}}$ or singular value $\sigma_{\scalebox{.7}{\{N,P\}=\{1,1\}}}$ of a single subarray system with a single path. The spectral efficiencies are examined with total transmit power of $20\ \mathrm{dBm}$ at different heights. Considering
Fig.~\ref{arraypattern}, Fig.~\ref{singreflSISOP20}, and Fig.~\ref{singlosSISO}, it can be found that when there is sufficient power and the directions of paths are aligned with the main lobe of the respective beam pattern, the spectral efficiency can be maximized. Meanwhile, the singular value spread is expected to be with the smallest distance regarding less of the oscillation phenomenon.

An oscillation phenomenon due to interference is observed. Invoking the narrow-band assumption, i.e. assuming that the symbol duration is longer than the delay spread, we explain the oscillation phenomenon as follows: the waves of the same transmitted symbol (first element in $\mathbf{z}$ as an example) are passing along 2 paths with different lengths. Therefore, they are superimposed with varying phase differences by varying antenna height $h$. This causes a periodic change between constructive and destructive interference. This interference phenomenon is influenced by the pathloss differences and the array gain differences. Only if the amplitudes of the same signal from the LoS path and the reflected path are comparable, strong oscillations in the magnitude of the singular values and the capacities occur (e.g., in low height region). Otherwise, the path with more power will dominate the received power of the respective pattern. 

The spatial frequency $f_h(h)$ of the oscillation w.r.t. height can be calculated by the length difference of the two paths. A length difference in the order of $\lambda$ is capable of leading several oscillating periods. The spatial frequency $f_h(h)$ is a function of $h$ as $f_h(h)=4\pi\sqrt{h^2+(\frac{D}{2})^2}/(h\lambda)$ due to the fact that the length of the second path is changing with $h$.

In Fig.~\ref{singreflSISOP20}, an interesting phenomenon is found that the spectral efficiency gains of non-orthogonal beam pairs (e.g., $\{\beta_1, \beta_2\}=\{0, \pi/8\}$, $\{\beta_1, \beta_2\}=\{0, 3\pi/8\}$) saturate at values higher than one at high height range, where only LoS path is dominating the system performance. This inspires a possible future work on using different array pattern with the same path. We note that this gain is not coming from the spatial multiplexing (as $N_\mathrm{s}=1$) offered by the channel, but a complex effective array gain of non-orthogonal patterns.

\subsection{Performance of a Multi-Subarray MIMO system with Large Subarray Separation}

\begin{figure}
\subfloat[\small $P_\mathrm{T}=20$~dBm]{
\includegraphics[width=.47\textwidth]{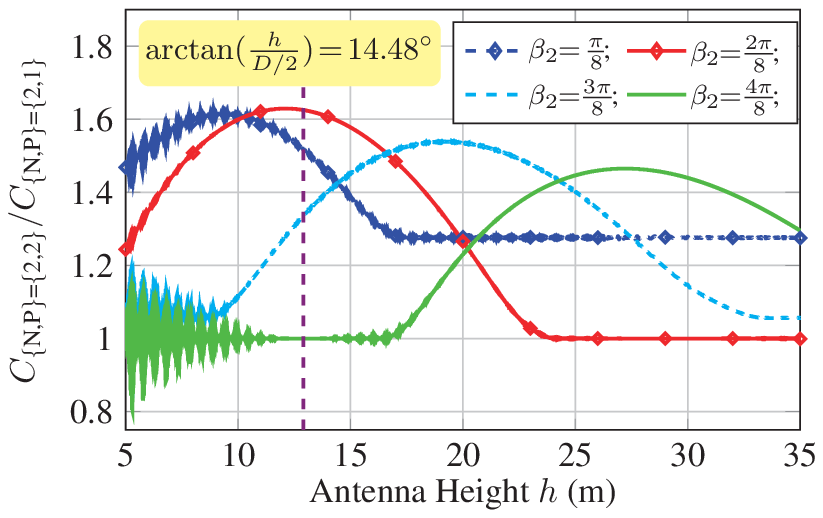}}
\hspace{3mm}
\subfloat[\small $P_\mathrm{T}=5$~dBm]{
\vspace{0.86mm}
\includegraphics[width=.47\textwidth]{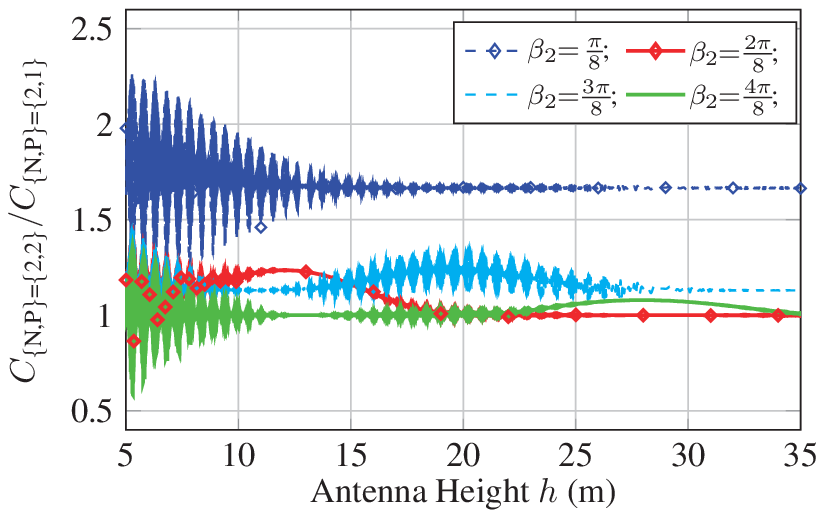}}
\vspace{-3mm}
\caption{\small Normalized capacities of a 2-path 2-subarray MIMO system w.r.t the capacity of an optimally arranged LoS MIMO system ($N=2$, $M=8$) where $C_{\protect\scalebox{.7}{\{N,P\}=\{2,1\}}}=2\cdot C_{\protect\scalebox{.7}{\{N,P\}=\{1,1\}}}$.} \label{Capacity_2path_2subarray}
\vspace*{-7mm}
\end{figure}

\begin{figure}
\subfloat[\small $\{\beta_1, \beta_2\}=\{0, \pi/8\}$]{
\includegraphics[width=.47\textwidth]{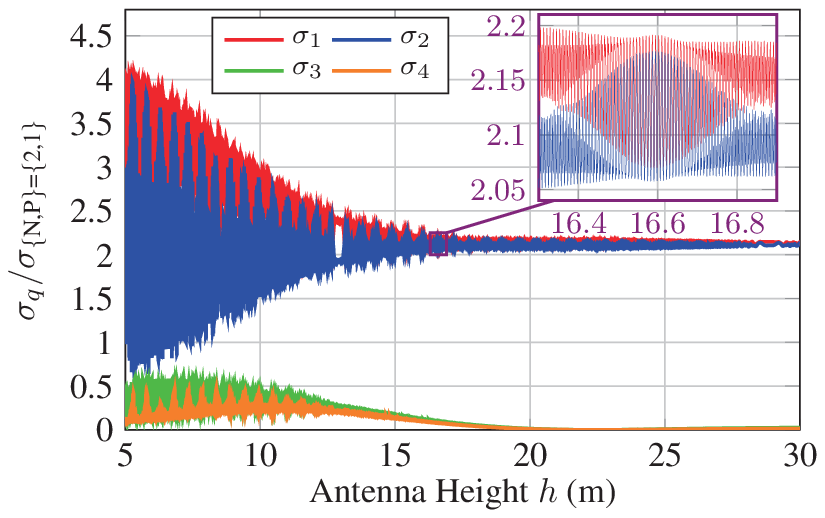}}
\hspace{3mm}
\subfloat[\small $\{\beta_1, \beta_2\}=\{0, 2\pi/8\}$]{
\vspace{0.87mm}
\includegraphics[width=.47\textwidth]{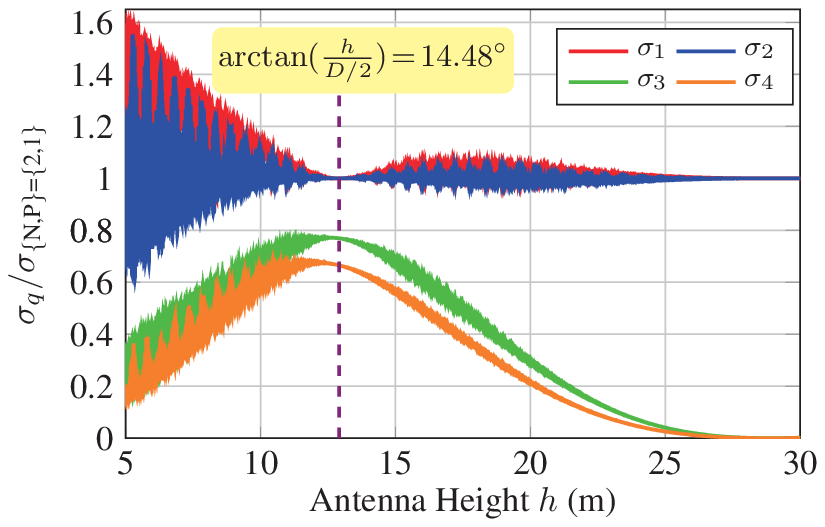}}
\vspace{-3mm}
\caption{\small Normalized singular values of a 2-path 2-subarray MIMO system w.r.t the singular value of an optimally arranged LoS MIMO system ($N=2$, $M=8$) where $\sigma_{\protect\scalebox{.7}{\{N,P\}=\{2,1\}}}=\sqrt{2} \cdot \sigma_{\protect\scalebox{.7}{\{N,P\}=\{1,1\}}}$.} \label{singular_2path_2subarray}
\vspace*{-7mm}
\end{figure}

Fig.~\ref{Capacity_2path_2subarray} and Fig.~\ref{singular_2path_2subarray} compare the spectral efficiencies and singular values of a 2-path channel when different beam patterns are used for a 2-subarray hybrid MIMO system, $N=2$. All the values are normalized w.r.t. the spectral efficiency $C_{\scalebox{.7}{\{N,P\}=\{2,1\}}}$ or the singular value $\sigma_{\scalebox{.7}{\{N,P\}=\{2,1\}}}$ of a two subarray system with a single path (LoS MIMO). From LoS MIMO theories, we know that the capacities for LoS MIMO systems with optimal arrangements are $N$ times larger than the capacity of a single subarray LoS system, then $C_{\scalebox{.7}{\{N,P\}=\{2,1\}}}=2\cdot C_{\scalebox{.7}{\{N,P\}=\{1,1\}}}$. Meanwhile, the singular values have a unique value of $\sigma_{\scalebox{.7}{\{N,P\}=\{2,1\}}}$ and $\sigma_{\scalebox{.7}{\{N,P\}=\{2,1\}}}=\sqrt{2} \cdot \sigma_{\scalebox{.7}{\{N,P\}=\{1,1\}}}$. 

Comparing Fig.~\ref{Capacity_2path_2subarray}(a) with Fig.~\ref{singreflSISOP20}, the gain brought by
multipath is almost the same, even if the antenna arrangements are just optimized for one
particular direction. This can be explained by the robustness of the single path MIMO
gain when introducing displacement errors like translation and rotation \cite{Larsson}.
Therefore, multi-subarray MIMO systems with large subarray separation provide a great
potential in approximate linearly scaling the throughput of a single subarray system in
a multipath environment. From Fig.~\ref{singular_2path_2subarray}, it is observed that the singular values are grouped with corresponding paths. Singular values within one group are closer to each other than the others. The group of singular values also show that the spectral efficiency can be scaled almost linearly with the number of subarrays that are large spaced, as only a single curve is found for each path in single subarray scenarios, Fig.~\ref{singlosSISO}.

The oscillation phenomenon discussed in Section~\ref{SISO_result} is also observed with multi-subarray MIMO systems. Additionally, besides the decreasing dynamic range of the spectral efficiency, beats of the oscillation frequency are also observed. The height difference of the two subarrays are causing a spatial frequency offset on $f_h(h)$. Therefore, when the interference of the signals from two subarrays are simultaneously in a constructive/destructive phase, this leads to large dynamic periods. Otherwise, if the constructive/destructive phases of the two subarrays are anti-phase, the spectral efficiencies are in low dynamic periods. The same phenomenon is also
observed in Fig.~\ref{singular_2path_2subarray}. Taking $\sigma_1$ and $\sigma_2$ as an example ($\sigma_1\geq \sigma_2$), the dynamic of the $\sigma_1$ curve is changing simultaneously with $\sigma_2$ curve.

With additional simulations using less transmit power Fig.~\ref{Capacity_2path_2subarray}(b), it is also observed that the oscillation is getting more severe as the available transmit power getting lower. Considering the waterfilling algorithm, if the fill-in water has low amount, the dynamic of container bottoms is causing the sensitiveness of the water level. Therefore, the dynamic of spectral efficiency is getting larger with larger dynamic on singular values and less transmit power.

\begin{figure}
\centering
\begin{minipage}{.47\textwidth}
\centering
\includegraphics[width=\linewidth]{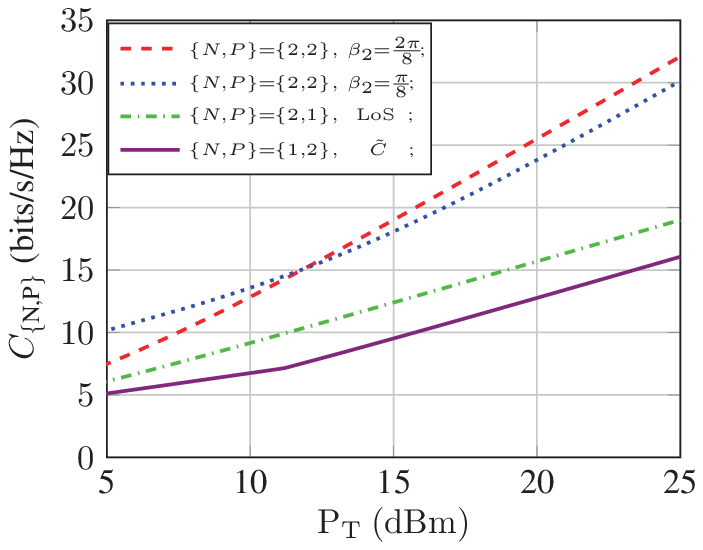}
\center
\vspace{-2mm}
\caption{\small Spectral efficiencies of hybrid MIMO systems when $h=\frac{D}{2}\tan(14.48^\circ)$.}
\label{fig:capacity_total_power}
\end{minipage}
\hspace{3mm}
\begin{minipage}{.47\textwidth}
\centering
\vspace{-1.5mm}
\includegraphics[width=1.03\linewidth]{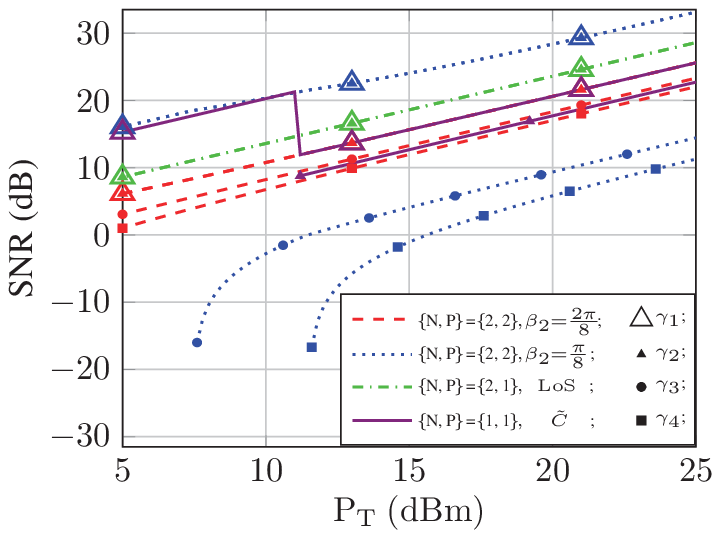}
\center
\vspace{-2mm}
\caption{\small SNRs of the subchannels of hybrid MIMO systems when $h=\frac{D}{2}\tan(14.48^\circ)$.}
\label{fig:snrs_total_power}
\vspace*{-3mm}
\end{minipage}
\end{figure}

Fig.~\ref{fig:capacity_total_power} and Fig.~\ref{fig:snrs_total_power} illustrate the variation of the capacity and the subchannel SNRs of different systems at different transmit power values but of the same height. The red-dashed line and the blue-dotted line indicate the achieved spectral efficiency/SNRs of a 2-path 2-subarray hybrid MIMO system when different beam pairs (value of $\beta_2$) are selected. Meanwhile the green-dashed-dotted and violet-solid line are the achievable spectral efficiency/SNRs of a conventional 2-subarray LoS MIMO system and a 2-path 1-subarray system. For simplification, the evaluations are carried out for height $h=\frac{D}{2}\tan(14.48^\circ)$, where the reflected path aligns with the main lobe of beam $\beta_2=2\pi/8$ and a null point of beam $\beta_1=0$. In Fig.~\ref{fig:capacity_total_power}, the achievable spectral efficiency of 2-path 2-subarray hybrid MIMO systems are almost doubling the values achieved by a single subarray system and are much higher than the values of the LoS only MIMO system.

In Fig.~\ref{fig:snrs_total_power}, it can be found that, at this given height, the SNRs for the first two
subchannels are almost the same for a 2-path 2-subarray MIMO system with large subarray separation. For the orthogonal beam pair $\{\beta_1, \beta_2\}=\{0, 2\pi/8\}$, the SNRs of the subchannels are closer to each other than the ones of
non-orthogonal beam pair, e.g., $\{\beta_1, \beta_2\}=\{0, \pi/8\}$. For non-orthogonal beam
pair $\{\beta_1, \beta_2\}=\{0, \pi/8\}$, only the first two good subchannels are used for
transmission with low transmit power amount, e.g., $P_\mathrm{T}=5 \mathrm{dBm}$. Meanwhile, the
first two subchannels of the non-orthogonal beams have higher SNRs than the first two
of the orthogonal ones. This gain is contributed by the complex array gains which we discussed earlier. Meanwhile, it can be found that at low transmit power, this gain is capable of providing higher spectral efficiencies
as in Fig.~\ref{Capacity_2path_2subarray}(b) and Fig.~\ref{fig:capacity_total_power}. However, the better aligned beam pair ($\{\beta_1, \beta_2\}=\{0, 2\pi/8\}$) achieves higher spectral efficiencies at high transmit power range as the cases for wireless backhaul systems.

For the 1-path (LoS) 2-subarray MIMO system and the 2-path 1-subarray MIMO system, the number of subchannels is reduced to two ($N_\mathrm{s}=2$, less spatial multiplexing) as shown in Fig.~\ref{fig:snrs_total_power}. When applying the waterfilling algorithm with the same amount of fill-in 'water' to the LoS MIMO system ($N_\mathrm{s}=2$), higher SNRs (green-dashed-dotted) are observed with less spectral efficiency in comparison with a 2-path 2-subarray hybrid MIMO system ($N_\mathrm{s}=4$). However, SNRs of the 2-paths 1-subarray hybrid MIMO are almost aligning with the SNRs of the 2-path 2-subarray system. The sudden change/apparent of SNRs at $P_{\mathrm{T}}$ value around $12~\mathrm{dBm}$ is due to a change of the selected beam pair (from $\{\beta_1, \beta_2\}=\{0, \pi/8\}$ to $\{\beta_1, \beta_2\}=\{0, 2\pi/8\}$).

\section{Conclusion}
\label{sec:conclusion}
In this work, a multi-subarray MIMO system design with large subarray separation is proposed for millimeter wave MIMO communication. Our work includes a multi-path channel model for such systems, and a hybrid beamforming architecture that achieves high spatial multiplexing. In comparison to state-of-the-art LoS MIMO based approaches, data rate increment of 50~$\%$ is observed by utility of just one additional path. Furthermore, in comparison with systems exploiting multiplexing gains between paths in a limited
scattering environment, an approximately linear scaling on the spectral efficiency is
observed. The spatial multiplexing gain of individual paths is determined by geometry
properties of the antenna arrangements and the path directions. It can also be found
that the geometry-relations/path-directions have a strong impact on the gains.
Furthermore, the number of active subchannels (spatial multiplexing) is also influenced by the available transmit
power. The subchannels are also compared for different analog beam patterns in this work. The proposed multi-subarray MIMO
system with large subarray separation shows a great potential in further increasing the
spectral efficiency with restricted path numbers/directions in applications like
wireless backhaul.

\appendix[Wave Models for Different Antenna Spacing]\label{appendix_1}

\begin{figure}[t]
\centering
\includegraphics[width=0.50\textwidth]{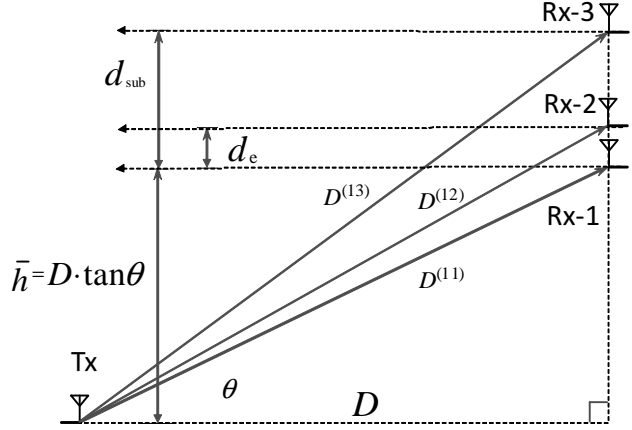}
\center
\vspace*{-3mm}
\caption{\small System sketch for a source transmit points and three receive antennas.}
\label{fig:wave_model}
\vspace*{-8mm}
\end{figure}

Let us consider a scenario with a source transmit point and three receive antenna elements as shown in Fig.~\ref{fig:wave_model}. The transmitter is $D$ meters away from the line connecting the receive antennas and the first receive antenna (Rx-1) is $\bar{h}$ meters away from the projected point. Let's set the phase center of the receive antennas at Rx-1. For simplification, we define $d_{i}$ indicating the inter spacing between Rx-$i$ and Rx-$1$. The second antenna element (Rx-2) is separated by a distance $d_{i}=d_{\mathrm{e}}$ in the order of wavelength $\lambda$, while the third one (Rx-3) is separated from Rx-1 by a larger distance $d_{i}=d_{\mathrm{sub}}\gg \lambda$.

For simplification of later calculation, we also include the elevation angle $\theta$ of the source point w.r.t. Rx-1. $D^{(11)}$, $D^{(12)}$ and $D^{(13)}$ are the distances between source point and respective receive antenna, where $D^{(11)}=\sqrt{{\bar{h}}^2+D^2}$. We assume that all geometry relations $D$, $D^{(11)}$, $D^{(12)}$ and $D^{(13)}$ are much larger than $d_{\mathrm{sub}}$,  $\{D$, $D^{(11)}$, $D^{(12)}$, $D^{(13)}\} \gg d_{\mathrm{sub}} \gg \lambda$.

Therefore, the relative phase $\phi_i$ of the wavefront arriving at Rx-$i$ can be expressed as
\begin{eqnarray}
e^{j\phi_i} &=& e^{-j\frac{2\pi}{\lambda} (D^{(1i)}-D^{(11)})} \\ 
 &=& e^{-j\frac{2\pi}{\lambda} \big[\sqrt{(\bar{h}+d_i)^2+D^2}-\sqrt{{\bar{h}}^2+D^2}\big]} \\
 &=& e^{-j\frac{2\pi\sqrt{{\bar{h}}^2+D^2}}{\lambda} \big[(1+\frac{2\bar{h}d_i}{{\bar{h}}^2+D^2}+\frac{d_i^2}{{\bar{h}}^2+D^2})^{1/2}-1\big]} \\
 &=& e^{-j\frac{2\pi\sqrt{{\bar{h}}^2+D^2}}{\lambda} \big[(1+2\sin\theta \frac{d_i}{\sqrt{{\bar{h}}^2+D^2}}+(\frac{d_i}{\sqrt{{\bar{h}}^2+D^2}})^2)^{1/2}-1\big]}.
\end{eqnarray}
Applying Taylor expansion to the term in bracket of the experiential part, the relative phase $\phi_i$ can be written as
\begin{equation}
\phi_i = -\frac{2\pi\sqrt{{\bar{h}}^2+D^2}}{\lambda} \big[\sin\theta \frac{d_i}{\sqrt{{\bar{h}}^2+D^2}}+(\frac{1-\sin^2\theta}{2})(\frac{d_i}{\sqrt{{\bar{h}}^2+D^2}})^2+\dots\big], \label{equ:wave_model}
\end{equation}
where we keep the expansion up to the second order of $d_i/\sqrt{{\bar{h}}^2+D^2}$.

To simplify the discussion later, we define a distance ratio $a_i$ as $a_i\triangleq d_i/\sqrt{\lambda \sqrt{{\bar{h}}^2+D^2}}=d_i/\sqrt{\lambda D^{(11)}}$. By replacing the $d_i$ in the second order term by $a_i$, Equ.~(\ref{equ:wave_model}) can be written in form of
\begin{equation}
\phi_i = -2\pi\big[\sin\theta \frac{d_i}{\lambda} +(\frac{1-\sin^2\theta}{2})a_i^2+\dots\big], \label{equ:wave_model_ratio}
\end{equation}

As for Rx-2, antenna spacing $d_i=d_{\mathrm{e}}$ is in order of $\lambda$. Then $a_i= d_{\mathrm{e}}/\sqrt{\lambda D^{(11)}}$ has property $a_i\ll 1$, as used in \cite{Balanis2005} (e.g., if $d_{\mathrm{e}}=\lambda$, $a_i=\sqrt{\lambda/D^{(11)}}$). Only the first term in Equ.~(\ref{equ:wave_model_ratio}) has significant contribution and $\phi_i$ becomes
\begin{equation}
\phi_i \approx -\frac{2\pi \sin\theta d_{\mathrm{e}}}{\lambda}, \label{equ:wave_model_2}
\end{equation}
which is a plane wave model and proves that the approximation to planar wave model is applicable for antenna elements inside subarrays.

However, for much larger antenna spacing (e.g., Rx-3) $d_i=d_{\mathrm{sub}}$ in the order of $\sqrt{\lambda D^{(11)}}$, the ratio $a_i$ is no longer negligible $a_i\not\ll 1$ (e.g., $d_{\mathrm{sub}}=\sqrt{\lambda D^{(11)}}$, $a_i=1$) and Equ.~(\ref{equ:wave_model_ratio}) becomes
\begin{equation}
\phi_i \approx -2\pi \Big(\frac{ \sin\theta \cdot d_{\mathrm{3}}}{\lambda}+ \frac{1-\sin^2\theta}{2}\cdot a_i^2\Big), \label{equ:wave_model_3}
\end{equation}
which is not fitting to the planar wave model. In this case, the spherical wave model should be kept for antennas (subarrays) with large antenna (subarray) separation.

\bibliographystyle{ieeetran}
\bibliography{reference}

\begin{thebibliography}{10}
\providecommand{\url}[1]{#1}
\csname url@samestyle\endcsname
\providecommand{\newblock}{\relax}
\providecommand{\bibinfo}[2]{#2}
\providecommand{\BIBentrySTDinterwordspacing}{\spaceskip=0pt\relax}
\providecommand{\BIBentryALTinterwordstretchfactor}{4}
\providecommand{\BIBentryALTinterwordspacing}{\spaceskip=\fontdimen2\font plus
\BIBentryALTinterwordstretchfactor\fontdimen3\font minus
  \fontdimen4\font\relax}
\providecommand{\BIBforeignlanguage}[2]{{%
\expandafter\ifx\csname l@#1\endcsname\relax
\typeout{** WARNING: IEEEtran.bst: No hyphenation pattern has been}%
\typeout{** loaded for the language `#1'. Using the pattern for}%
\typeout{** the default language instead.}%
\else
\language=\csname l@#1\endcsname
\fi
#2}}
\providecommand{\BIBdecl}{\relax}
\BIBdecl

\bibitem{GF_2020}
G.~Fettweis, ``{LTE: The Move to Global Cellular Broadband},'' \emph{{Intel
  Technical Journal, special issue on LTE}}, vol.~18, pp. 7--10, Feb 2014.

\bibitem{Marzetta_massiveMIMO}
T.~L.Marzetta, ``Noncooperative cellular wireless with unlimited numbers of
  base station antennas,'' \emph{IEEE Trans. Wireless Comm.}, vol.~9, no.~11,
  p. 3590–3600, Nov. 2010.

\bibitem{Song_GC2015}
X.~Song, C.~Jans, L.~Landau, D.~Cvetkovski, and G.~Fettweis, ``{A 60GHz LOS
  MIMO Backhaul Design Combining Spatial Multiplexing and Beamforming for a
  100Gbps Throughput},'' in \emph{{Proceedings of the IEEE Global
  Communications Conference}}, Dec 2015.

\bibitem{Larsson}
P.~Larsson, ``{Lattice Array Receiver and Sender for Spatially Orthonormal MIMO
  Communication},'' in \emph{{Proceedings of the IEEE 61st Vehicular Technology
  Conference}}, vol.~1, May 2005, pp. 192--196.

\bibitem{Haustein2003}
T.~Haustein and U.~Kruger, ``{Smart Geometrical Antenna Design Exploiting the
  LOS Component to Enhance a MIMO System Based on Rayleigh-fading in Indoor
  Scenarios},'' in \emph{Proceedings of the 14th IEEE Proceedings on Personal,
  Indoor and Mobile Radio Communications}, vol.~2, Sept 2003, pp. 1144--1148.

\bibitem{BohagenConstructionCapacity}
F.~Bohagen, P.~Orten, and G.~Oien, ``{Construction and Capacity Analysis of
  High-rank Line-of-sight MIMO Channels},'' in \emph{Proceedings of the IEEE
  Wireless Communications and Networking Conference}, vol.~1, March 2005, pp.
  432--437.

\bibitem{BohagenOptimalDesignUPA}
------, ``{Optimal Design of Uniform Planar Antenna Arrays for Strong
  Line-of-Sight MIMO Channels},'' in \emph{Proceedings of the IEEE 7th Workshop
  on Signal Processing Advances in Wireless Communications}, July 2006, pp.
  1--5.

\bibitem{ChunhuiZhou}
C.~Zhou, X.~Chen, X.~Zhang, S.~Zhou, M.~Zhao, and J.~Wang, ``{Antenna Array
  Design for LOS-MIMO and Gigabit Ethernet Switch-Based Gbps Radio System},''
  \emph{{International Journal of Antennas and Propagation}}, 2012.

\bibitem{Song_2015}
X.~Song and G.~Fettweis, ``{On Spatial Multiplexing of Strong Line-of-Sight
  MIMO With 3D Antenna Arrangements},'' \emph{{IEEE Wireless Communications
  Letters}}, vol.~4, no.~4, pp. 393--396, Aug 2015.

\bibitem{Telatar99}
E.~Telatar, ``{Capacity of Multi-antenna Gaussian Channels},'' \emph{European
  Transactions on Telecommunications}, vol.~10, no.~6, pp. 585--595, 1999.

\bibitem{Foschini1996}
G.~J. Foschini, ``{Layered space--time architecture for wireless communication
  in a fading environment when using multiple antennas },'' \emph{{Bell Labs.
  Techn. J.}}, vol.~1, pp. 41 -- 59, 1996.

\bibitem{Hybrid_switching_05}
X.~Zhang, A.~F. Molisch, and S.-Y. Kung, ``{Variable-Phase-Shift-Based
  RF-Baseband Codesign for MIMO Antenna Selection},'' \emph{IEEE Transactions
  on Signal Processing}, vol.~53, no.~11, pp. 4091--4103, Nov 2005.

\bibitem{Veen_Hybrid_10}
V.~Venkateswaran and A.~J. van~der Veen, ``{Analog Beamforming in MIMO
  Communications With Phase Shift Networks and Online Channel Estimation},''
  \emph{IEEE Transactions on Signal Processing}, vol.~58, no.~8, pp.
  4131--4143, Aug 2010.

\bibitem{Heath_Hybrid_12}
O.~E. Ayach, R.~W. Heath, S.~Abu-Surra, S.~Rajagopal, and Z.~Pi, ``{Low
  Complexity Precoding for Large Millimeter Wave MIMO Systems},'' in
  \emph{{Proceedings of IEEE International Conference on Communications}}, June
  2012, pp. 3724--3729.

\bibitem{Pu2015}
X.~Pu, S.~Shao, K.~Deng, and Y.~Tang, ``{Analysis of the Capacity Statistics
  for 2 $\times$ 2 3D MIMO Channels in Short-Range Communications},''
  \emph{IEEE Communications Letters,}, vol.~19, no.~2, pp. 219--222, Feb 2015.

\bibitem{Mailloux2005}
R.~Mailloux, \emph{Phased Array Antenna Handbook}.\hskip 1em plus 0.5em minus
  0.4em\relax Artech House, 2005.

\bibitem{Sayeed_07}
A.~M. Sayeed and V.~Raghavan, ``Maximizing mimo capacity in sparse multipath
  with reconfigurable antenna arrays,'' \emph{IEEE Journal of Selected Topics
  in Signal Processing}, vol.~1, no.~1, pp. 156--166, June 2007.

\bibitem{Heath_2014_channel_model}
O.~E. Ayach, S.~Rajagopal, S.~Abu-Surra, Z.~Pi, and R.~W. Heath, ``{Spatially
  Sparse Precoding in Millimeter Wave MIMO Systems},'' \emph{{IEEE Transactions
  on Wireless Communications}}, vol.~13, no.~3, pp. 1499--1513, March 2014.

\bibitem{Balanis2005}
C.~A. Balanis, \emph{{Antenna Theory: Analysis and Design, 3rd Edition}}.\hskip
  1em plus 0.5em minus 0.4em\relax Wiley-Interscience, Apr. 2005.

\bibitem{rappaport}
T.~Rappaport, \emph{Wireless Communications: Principles and Practice},
  2nd~ed.\hskip 1em plus 0.5em minus 0.4em\relax Upper Saddle River, NJ, USA:
  Prentice Hall PTR, 2001.

\bibitem{Loan2000}
C.~F. Loan, ``{The Ubiquitous Kronecker Product },'' \emph{{Journal of
  Computational and Applied Mathematics}}, vol. 123, no. 1–2, pp. 85 -- 100,
  2000.

\bibitem{rf}
A.~Hajimiri, H.~Hashemi, A.~Natarajan, X.~Guan, and A.~Komijani, ``Integrated
  phased array systems in silicon,'' \emph{Proceedings of the IEEE}, vol.~93,
  no.~9, pp. 1637--1655, Sept 2005.

\bibitem{Heath_2016_waterfilling}
A.~Alkhateeb and R.~W. Heath, ``Frequency selective hybrid precoding for
  limited feedback millimeter wave systems,'' \emph{IEEE Transactions on
  Communications}, vol.~64, no.~5, pp. 1801--1818, May 2016.

\bibitem{CoverThomas}
T.~M. Cover and J.~A. Thomas, \emph{Elements of Information Theory},
  2nd~ed.\hskip 1em plus 0.5em minus 0.4em\relax John Wiley \& Sons, 2006.

\bibitem{concrete}
B.~Feitor, R.~Caldeirinha, T.~Fernandes, D.~Ferreira, and N.~Leonor,
  ``{Estimation of Dielectric Concrete Properties from Power Measurements at
  18.7 and 60 GHz},'' in \emph{Antennas and Propagation Conference (LAPC), 2011
  Loughborough}, Nov 2011, pp. 1--5.

\bibitem{IEEE-Std-802.11ad}
``{IEEE Standard for Information technology--Telecommunications and information
  exchange between systems--Local and metropolitan area networks--Specific
  requirements-Part 11: Wireless LAN Medium Access Control (MAC) and Physical
  Layer (PHY) Specifications Amendment 3: Enhancements for Very High Throughput
  in the 60 GHz Band},'' \emph{IEEE Std 802.11ad-2012 (Amendment to IEEE Std
  802.11-2012, as amended by IEEE Std 802.11ae-2012 and IEEE Std
  802.11aa-2012)}, p. 443, Dec 2012.

\bibitem{FCCEIRP}
F.~C. Commission, \emph{{Part 15 of the Commission's Rules Regarding
  Operation in the 57-64 GHz Band, ET Docket No. 07-113 and RM-11104}}, August
  2013.

\end{thebibliography}

\end{document}